\documentclass[preprint,showpacs,preprintnumbers,superscriptaddress,nofootinbib]{revtex4}

\usepackage{graphicx}
\usepackage{dcolumn}
\usepackage{bm}
\usepackage{amsmath}
\usepackage{amssymb}
\usepackage{url}
\usepackage[normalem]{ulem}
\usepackage{xcolor}

\newcommand{\beq}{\begin{equation}}
\newcommand{\eeq}{\end{equation}}
\newcommand{\bea}{\begin{eqnarray}}
\newcommand{\eea}{\end{eqnarray}}
\newcommand{\real}{{\sf I}\kern-.12em{\sf R}}
\newcommand{\comp}{{\sf I}\kern-.50em{\sf C}}
\newcommand{\unity}{{\sf I}\kern-.54em{\sf 1}}

\def\spose#1{\hbox to 0pt{#1\hss}}
\def\ltapprox{\mathrel{\spose{\lower 3pt\hbox{$\mathchar"218$}}
 \raise 2.0pt\hbox{$\mathchar"13C$}}}

\begin{document}

\title{Hadron-jet correlations \\  
  in high-energy hadronic collisions at the LHC}
\author{Andr\`ee D. Bolognino}
\affiliation{Dipartimento di Fisica dell'Universit\`a della Calabria \\
I-87036 Arcavacata di Rende, Cosenza, Italy}
\affiliation{INFN - Gruppo collegato di Cosenza, I-87036 Arcavacata di Rende,
Cosenza, Italy}
\author{Francesco G. Celiberto}
\affiliation{Instituto de F{\' \i}sica Te{\' o}rica UAM/CSIC, Nicol{\'a}s Cabrera 15, 28049 Madrid, Spain}
\affiliation{Universidad Aut{\'o}noma de Madrid, 28049 Madrid, Spain}
\author{Dmitry Yu. Ivanov}
\affiliation{Sobolev Institute of Mathematics, 630090 Novosibirsk, Russia}
\affiliation{Novosibirsk State University, 630090 Novosibirsk, Russia}
\author{Mohammed M.A. Mohammed}
\affiliation{Dipartimento di Fisica dell'Universit\`a della Calabria \\
I-87036 Arcavacata di Rende, Cosenza, Italy}
\author{Alessandro Papa}
\affiliation{Dipartimento di Fisica dell'Universit\`a della Calabria \\
I-87036 Arcavacata di Rende, Cosenza, Italy}
\affiliation{INFN - Gruppo collegato di Cosenza, I-87036 Arcavacata di Rende,
Cosenza, Italy}

\date{\today}

\begin{abstract}
  The inclusive production at the LHC of a charged light hadron and of a
  jet, featuring a wide separation in rapidity, is suggested as a new probe
  process for the investigation of the BFKL mechanism of resummation of energy
  logarithms in the QCD perturbative series.
  We present some predictions, tailored on the CMS and CASTOR acceptances,
  for the cross section averaged over the azimuthal angle between the
  identified jet and hadron and for azimuthal correlations.
\end{abstract}

\maketitle

\section{Introduction}
\label{introd}

The LHC record energy, as well as the good resolution in azimuthal angles of
the particle detectors, offer a unique opportunity to test a wide class
of predictions of perturbative QCD. These include the so called Mueller-Navelet
jet production~\cite{Mueller:1986ey}, {\it i.e.} the inclusive production of
two jets featuring a large rapidity separation between them, for which
a wealth of theoretical analyses were produced in the last years~\cite{Colferai:2010wu,Angioni:2011wj,Caporale:2012ih,Ducloue:2013wmi,Ducloue:2013bva,Caporale:2013uva,Ducloue:2014koa,Caporale:2014gpa,Ducloue:2015jba,Caporale:2015uva,Celiberto:2015yba,Celiberto:2016ygs,Chachamis:2015crx,Caporale:2018qnm}, and the
somewhat related process where two identified, charged light hadrons well
separated in rapidity are inclusively
produced~\cite{Celiberto:2016hae,Celiberto:2017ptm}, instead of jets. 

A common feature to these two processes is that their high-energy behavior
is dominated by those final-state configurations where the produced particles
are strongly ordered in rapidity, the tagged objects (jets or identified
hadrons) being the two extrema in the rapidity tower, thus yielding a number
of energy logarithms growing with the number of produced particles. Such
energy logarithms are so large to compensate the smallness
of the coupling $\alpha_s$, so the perturbative series must be properly
resummed.

The theoretical framework for the resummation of energy logs for
these two processes, as well as for any semi-hard process in perturbative QCD,
is provided by the Balitsky--Fadin--Kuraev--Lipatov (BFKL) approach~\cite{BFKL},
whereby the resummation of all terms proportional to $(\alpha_s\ln(s))^n$,
the so called leading logarithmic approximation or LLA, and that of all
terms proportional to $\alpha_s(\alpha_s\ln(s))^n$, the next-to-leading
approximation or NLA, can be systematically carried out.
The bottom line of the BFKL formalism is that azimuthal coefficients
of the Fourier expansion of the cross section differential in the variables
of the tagged objects over the relative azimuthal angle take the very
simple form of a convolution between two impact factors, describing
the transition from each colliding proton to the respective 
final state tagged object, and a process-independent Green's function.
The BFKL Green's function obeys an integral equation, whose
kernel is known at the next-to-leading order (NLO) both for forward
scattering ({\it i.e.} for $t=0$ and color singlet in the
$t$-channel)~\cite{Fadin:1998py,Ciafaloni:1998gs} 
and for any fixed (not growing with energy)
momentum transfer $t$ and any possible two-gluon color state in the
$t$-channel~\cite{Fadin:1998jv,FG00,FF05}.

The impact factors for the proton to forward jet transition
(the so called ``jet vertices'') are known up to the NLO for several jet
selection algorithms~\cite{Bartels:2001ge,Bartels:2002yj,Caporale:2011cc,Ivanov:2012ms,Colferai:2015zfa}. The jet vertex, in its turn, can be expressed, within
leading-twist collinear factorization, as the convolution of the parton
distribution function (PDF) of the colliding proton, obeying the standard
DGLAP evolution~\cite{DGLAP}, with the hard process describing the transition
from the parton emitted by the proton to the forward jet in the final state.
Two such jet vertices must be convoluted with the BFKL Green's function to
theoretically describe the Mueller-Navelet jet production. The main aim is
to calculate cross sections and azimuthal angle
correlations~\cite{DelDuca:1993mn,Stirling:1994he} between the two measured
jets, {\it i.e.} average values of $\cos{(n \phi)}$, where $n$ is an integer
and $\phi$ is the angle in the azimuthal plane
between the direction of one jet and the direction opposite to the other jet,
and ratios of two such cosines~\cite{Vera:2006un,Vera:2007kn}. 

Also the impact factors for the proton to identified hadron transition
are known up to the NLO~\cite{hadrons} and can be expressed, within
leading-twist collinear factorization, as the convolution of the parton
distribution function (PDF) of the colliding proton with the hard process
describing the transition from the parton emitted by the proton to a
final-state parton and with the fragmentation function (FF) for that parton
to the desired hadron. Two such  hadron  vertices must be convoluted with the BFKL
Green's function to theoretically describe the above-mentioned inclusive
hadron-hadron production and finally get predictions for cross sections and
azimuthal angle correlations, similarly to the case of jets.

Within the same formalism, other interesting processes have been
proposed as a testfield for BFKL dynamics at the LHC, namely the inclusive
production of three or four jets, well separated in rapidity
from each other~\cite{Caporale:2015vya,Caporale:2016soq,Caporale:2016zkc,Caporale:2015int,Caporale:2016xku},
the inclusive detection of two heavy quark-antiquark pairs, separated in
rapidity, in the collision of two real (or quasi-real)
photons~\cite{Celiberto:2017nyx}, and the inclusive tag of a forward $J/\Psi$-meson and a very backward jet at the LHC~\cite{Boussarie:2017oae}. 

On the experimental side the situation is as follows: the CMS
Collaboration~\cite{Khachatryan:2016udy} has presented the first measurements
of the azimuthal correlation of the Mueller-Navelet jets at $\sqrt{s}=7$~{TeV},
but further experimental studies of the Mueller-Navelet jets are expected
at higher LHC energies and larger rapidity intervals, including also the
effects of using asymmetrical cuts for the jet transverse momenta. No
experimental analyses have yet appeared on azimuthal correlation between
two rapidity-separated identified light hadrons. The reason for that could
be that events with identified hadrons in the final state, carrying
transverse momenta of the order of, say, 5~GeV or larger, fall into the class
of what experimentalists call ``minimum bias events'', which represent the
main background in high-luminosity runs at a collider. They would be better
studied in low-luminosity, dedicated, runs.

\begin{figure}[t]
 \centering
 \includegraphics[scale=0.5]{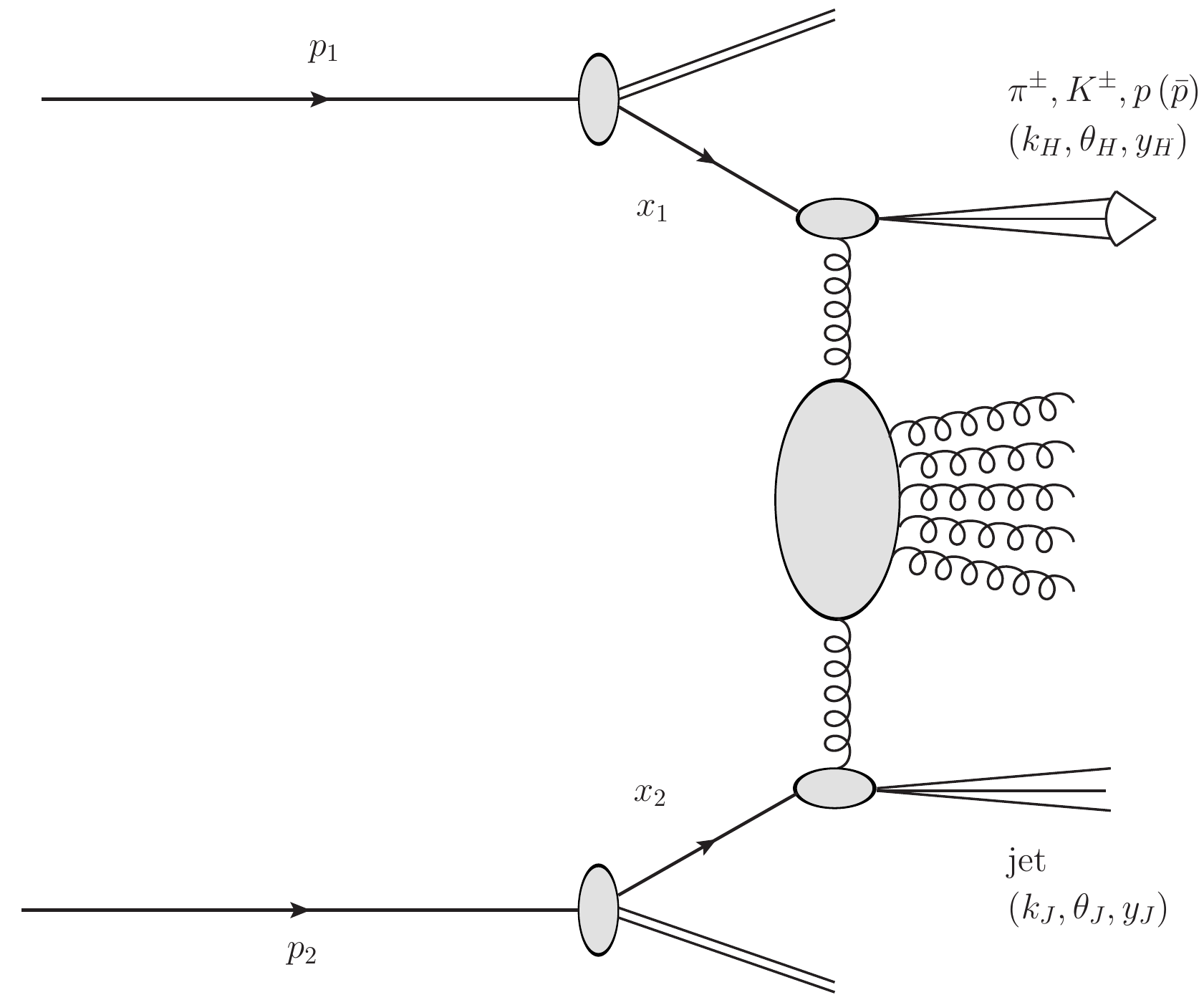}
 \caption[]
         {Inclusive hadroproduction of a charged light hadron and of a jet.}
           \label{fig:hadron-jet}
 \end{figure}

In this paper we want to introduce a new process which could serve as a
probe of BFKL dynamics: the inclusive hadron-jet production in proton-proton
collisions, 
\begin{eqnarray}
\label{process}
{\rm proton}(p_1) + {\rm proton}(p_2) 
\to 
{\rm hadron}(k_H, y_H) + {\rm X} + {\rm jet}(k_J, y_J) \;,
\end{eqnarray}
when a charged light hadron: $\pi^{\pm}, K^{\pm}, p \left(\bar p\right)$ and a
jet with high transverse momenta, separated by a large interval of rapidity,
are produced together with an undetected hadronic system X
(see Fig.~\ref{fig:hadron-jet} for a schematic view). The
process~(\ref{process}) has many common features with the inclusive
$J/\Psi$-meson plus backward jet production, considered recently in
Ref.~\cite{Boussarie:2017oae}. From the experimental side, the detection of
the $J/\Psi$-meson looks rather appealing. But, from the theory side, there are
more uncertainties in this case in comparison to our proposal. The
$J/\Psi$-meson production impact factor was considered in LO; moreover,
several production mechanisms in the frame of NRQCD were discussed. Instead,
the light hadron impact factor is well defined in collinear factorization and
it is known in NLO. Previous experience in BFKL calculations for various
processes at LHC shows that the account of NLO corrections to the impact
factors leads both to a considerable change of predictions and to a big
reduction of the theoretical uncertainties.

The theoretical task to build predictions for cross section and azimuthal
correlations for our process is embarrassingly simple: one should simply
replace one of the two jet impact factor entering the Mueller-Navelet formulas
with the vertex for the proton-to-hadron transition. From the theoretical point
of view, this process is definitely an easy target, since all the needed
building blocks are available, with NLO accuracy.

Yet, we believe that there are some good reasons for building numerical
predictions for this process and submitting them to the attention of both
experimentalists and theorists:

\begin{itemize}

\item the BFKL resummation implies certain factorization structure for the
  predicted observables: the latter are calculated as a convolution of
  the universal BFKL Green's function with the process dependent impact factors,
  which resembles the factorization in Regge theory. It is important to test
  this picture experimentally, considering all possible processes for which the
  full NLO BFKL description is available.

\item In Refs.~\cite{Caporale:2014gpa,Celiberto:2015yba,Celiberto:2015mpa}
  it was discussed, in the context of Mueller-Navelet jet production, that
  using {\em asymmetric} cuts for the transverse momenta of the tagged jets
  suppresses the Born term, present only for back-to-back jets, thus enhancing
  the effects of the additional undetected hard gluon radiation and 
  making therefore more visible the impact of the BFKL resummation, with
  respect to the fixed-order (DGLAP) contribution. For the process we
  are considering here this kind asymmetry would be naturally imposed by the
  completely different nature of the two tagged objects: the identified jet
  should have transverse momentum not smaller than 20~GeV or so, whereas the
  minimum hadron transverse momentum can be as small as 5~GeV.

\item For the process under consideration only one hadron in the final state
  should be identified, instead of two as in the hadron-hadron inclusive
  production, the other identified object being a jet with a typically much
  larger transverse momentum. This should facilitate the mining of these
  events out of the minimum-bias ones.

\item From the theoretical point of view one can use this process to compare
  models for FFs or for jet algorithms, handling expressions which are {\em
  linear} in the corresponding functions and not quadratic as it would be,
  respectively, in the hadron-hadron and in the Mueller-Navelet jet case.

\end{itemize}

The summary of the paper is as follows: in Section~1 we present the theoretical
framework and sketch the derivation of our predictions; in Section~2 we
show and discuss the results of our numerical analysis; finally, in Section~3,
we draw our conclusions and give some outlook.

\section{Theoretical framework}
\label{theory}

The final state configuration of the inclusive process under consideration is
schematically represented in Fig.~\ref{fig:hadron-jet}, where a charged
light hadron $(k_H, y_H)$ and a jet $(k_J, y_J)$ are detected, featuring
a large rapidity separation, together with an undetected system of hadrons.
For the sake of definiteness, we will consider the case where the 
hadron rapidity $y_H$ is larger than the jet one  $y_J$, so that $Y\equiv
y_H-y_J$ is always positive. This implies that, for most of the considered
values of $Y$, the hadron is forward and the jet is backward.

The hadron and the jet are also required to possess large transverse momenta,
$\vec k_H^2\sim \vec k_J^2 \gg \Lambda^2_{\rm QCD}$.
The protons' momenta $p_1$ and $p_2$ are taken as Sudakov vectors satisfying
$p^2_1= p^2_2=0$ and $2 (p_1p_2) = s$,  so that the momenta of the final-state
objects can be decomposed as
\bea
k_H&=& x_H p_1+ \frac{\vec k_H^2}{x_H s}p_2+k_{H\perp} \ , \quad
k_{H\perp}^2=-\vec k_H^2 \ , \nonumber \\
k_J&=& x_J p_2+ \frac{\vec k_J^2}{x_J s}p_1+k_{J\perp} \ , \quad
k_{J\perp}^2=-\vec k_J^2 \ .
\label{sudakov}
\eea

In the center-of-mass system, the hadron/jet longitudinal momentum fractions 
$x_{H,J}$ are connected to the respective rapidities through the relations
$y_H=\frac{1}{2}\ln\frac{x_H^2 s}
{\vec k_H^2}$, and $y_J=\frac{1}{2}\ln\frac{\vec k_J^2}{x_J^2 s}$, 
so that $dy_H=\frac{dx_H}{x_H}$, $dy_J=-\frac{dx_J}{x_J}$,
and $Y=y_H-y_J=\ln\frac{x_Hx_J s}{|\vec k_H||\vec k_J|}$, here the
space part of the four-vector $p_{1\parallel}$ being taken positive.

In QCD collinear factorization the cross section of the
process~(\ref{process}) reads
\beq
\frac{d\sigma}{dx_Hdx_Jd^2k_Hd^2k_J}
=\sum_{r,s=q,{\bar q},g}\int_0^1 dx_1 \int_0^1 dx_2\ f_r\left(x_1,\mu_F\right)
\ f_s\left(x_2,\mu_F\right)
\frac{d{\hat\sigma}_{r,s}\left(\hat s,\mu_F\right)}
{dx_Hdx_Jd^2k_Hd^2k_J}\;,
\eeq
where the $r, s$ indices specify the parton types 
(quarks $q = u, d, s, c, b$;
antiquarks $\bar q = \bar u, \bar d, \bar s, \bar c, \bar b$; 
or gluon $g$), $f_{r,s}\left(x, \mu_F \right)$ denote the initial proton PDFs; 
$x_{1,2}$ are the longitudinal fractions of the partons involved in the hard
subprocess, while $\mu_F$ is the factorization scale;
$d\hat\sigma_{r,s}\left(\hat s \right)$ is
the partonic cross section and $\hat s \equiv x_1x_2s$ is the squared
center-of-mass energy of the parton-parton collision subprocess.

In the BFKL approach the cross section can be presented 
(see Ref.~\cite{Caporale:2012ih} for the details of the derivation)
as the Fourier sum of the azimuthal coefficients ${\cal C}_n$, 
having so
\beq
\frac{d\sigma}
{dy_Hdy_J\, d|\vec k_H| \, d|\vec k_J|d\phi_H d\phi_J}
=\frac{1}{(2\pi)^2}\left[{\cal C}_0+\sum_{n=1}^\infty  2\cos (n\phi )\,
{\cal C}_n\right]\, ,
\eeq
where $\phi=\phi_H-\phi_J-\pi$, with $\phi_{H,J}$ the hadron/jet 
azimuthal angles, while $y_{H,J}$ and $\vec k_{H,J}$ are their
rapidities and transverse momenta, respectively. 
The $\phi$-averaged cross section ${\cal C}_0$ 
and the other coefficients ${\cal C}_{n\neq 0}$ are given
by~\footnote{In Ref.~\cite{Celiberto:2017ptm}, on the last line of Eq.~(5),
  which is closely related to this formula for ${\cal C}_n$, it was mistakenly
  written $2\ln\left(\vec k_1^2 \vec k_2^2\right)$ instead of
  $\ln\left(\vec k_1^2 \vec k_2^2\right)$, although the numerical results
  presented there were obtained using the correct formula.}
\beq\nonumber
{\cal C}_n \equiv \int_0^{2\pi}d\phi_H\int_0^{2\pi}d\phi_J\,
\cos[n(\phi_H-\phi_J-\pi)] \,
\frac{d\sigma}{dy_Hdy_J\, d|\vec k_H| \, d|\vec k_J|d\phi_H d\phi_J}\;
\eeq
\beq\nonumber
= \frac{e^Y}{s}
\int_{-\infty}^{+\infty} d\nu \, \left(\frac{x_H x_J s}{s_0}
\right)^{\bar \alpha_s(\mu_R)\left\{\chi(n,\nu)+\bar\alpha_s(\mu_R)
\left[\bar\chi(n,\nu)+\frac{\beta_0}{8 N_c}\chi(n,\nu)\left[-\chi(n,\nu)
+\frac{10}{3}+2\ln\left(\frac{\mu_R^2}{\sqrt{\vec k_H^2\vec k_J^2}}\right)\right]\right]\right\}}
\eeq
\beq\nonumber
\times \alpha_s^2(\mu_R) c_H(n,\nu,|\vec k_H|, x_H)
[c_J(n,\nu,|\vec k_J|,x_J)]^*\,
\eeq
\beq\label{Cm}
\times \left\{1
+\alpha_s(\mu_R)\left[\frac{c_H^{(1)}(n,\nu,|\vec k_H|,
x_H)}{c_H(n,\nu,|\vec k_H|, x_H)}
+\left[\frac{c_J^{(1)}(n,\nu,|\vec k_J|, x_J)}{c_J(n,\nu,|\vec k_J|,
x_J)}\right]^*\right]\right.
\eeq
\beq\nonumber
\left. + \bar\alpha_s^2(\mu_R) \ln\left(\frac{x_H x_J s}{s_0}\right)
\frac{\beta_0}{4 N_c}\chi(n,\nu)f(\nu)\right\}\;.
\eeq
\\
Here $\bar \alpha_s(\mu_R) \equiv \alpha_s(\mu_R) N_c/\pi$, with
$N_c$ the number of colors,
\beq
\beta_0=\frac{11}{3} N_c - \frac{2}{3}n_f
\eeq
is the first coefficient of the QCD $\beta$-function, where $n_f$ is the number
of active flavors,
\beq
\chi\left(n,\nu\right)=2\psi\left(1\right)-\psi\left(\frac{n}{2}
+\frac{1}{2}+i\nu \right)-\psi\left(\frac{n}{2}+\frac{1}{2}-i\nu \right)
\eeq
is the leading-order (LO) BFKL characteristic function,
$c_H(n,\nu)$ is the LO forward hadron impact factor in the
$\nu$-repre\-sen\-ta\-tion,  given as an integral in the parton fraction $x$,
containing the PDFs of the gluon and of the different quark/antiquark flavors
in the proton, and the FFs of the detected hadron,
\bea
c_H(n,\nu,|\vec k_H|,x_H) &=& 2 \sqrt{\frac{C_F}{C_A}}
(\vec k_H^2)^{i\nu-1/2}\,\int_{x_H}^1\frac{dx}{x}
\left( \frac{x}{x_H}\right)
^{2 i\nu-1} 
\nonumber \\
\label{cH}
&\times&\left[\frac{C_A}{C_F}f_g(x)D_g^h\left(\frac{x_H}{x}\right)
+\sum_{r=q,\bar q}f_r(x)D_r^h\left(\frac{x_H}{x}\right)\right] \;,
\eea
$c_J(n,\nu)$ is the LO forward jet vertex in the $\nu$-repre\-sen\-ta\-tion,
\beq
\label{cJ}
c_J(n,\nu,|\vec k_J|,x_J)=2\sqrt{\frac{C_F}{C_A}}
(\vec k_J^{\,2})^{i\nu-1/2}\,\left(\frac{C_A}{C_F}f_g(x_J)
+\sum_{s=q,\bar q}f_s(x_J)\right)
\eeq
and the $f(\nu)$ function is defined by
\beq
\label{fnu}
i\frac{d}{d\nu}\ln\left(\frac{c_H}{[c_J]^*}\right)=2\left[f(\nu)
  -\ln\left(\sqrt{\vec k_H^2 \vec k_J^2}\right)\right] \;.
\eeq
The remaining objects are the hadron/jet NLO impact factor corrections in the
$\nu$-representation, $c_{H,J}^{(1)}(n,\nu,|\vec k_{H,J}|, x_{H,J})$, their
expressions being given in Eqs.~(4.58)-(4.65) of Ref.~\cite{hadrons}
and in Eq.~(36) of Ref.~\cite{Caporale:2012ih}, respectively.

\section{Results and Discussion}
\label{results}

\begin{figure}[b]
  \centering
  \includegraphics[scale=0.33,clip]{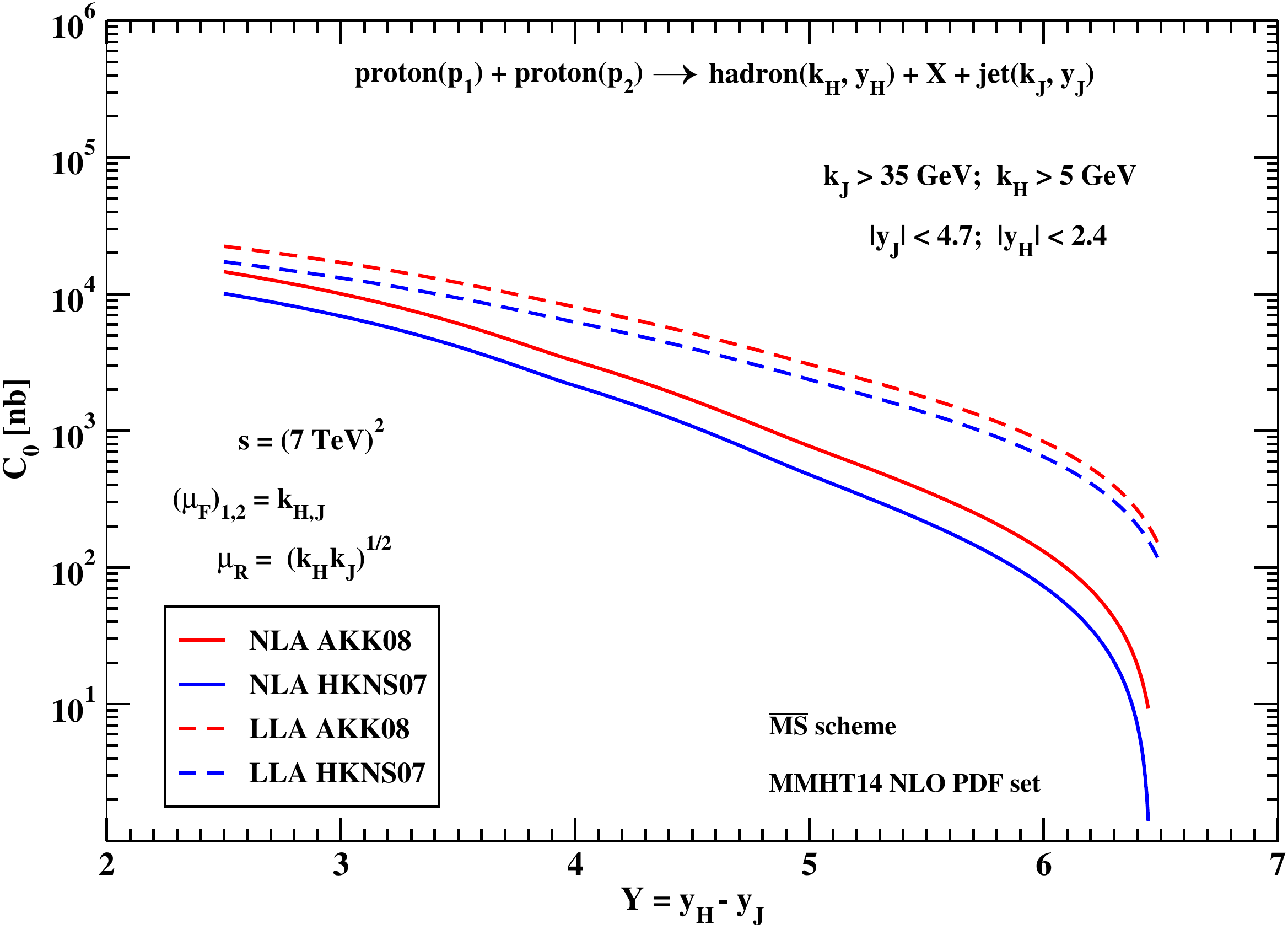}
  \includegraphics[scale=0.33,clip]{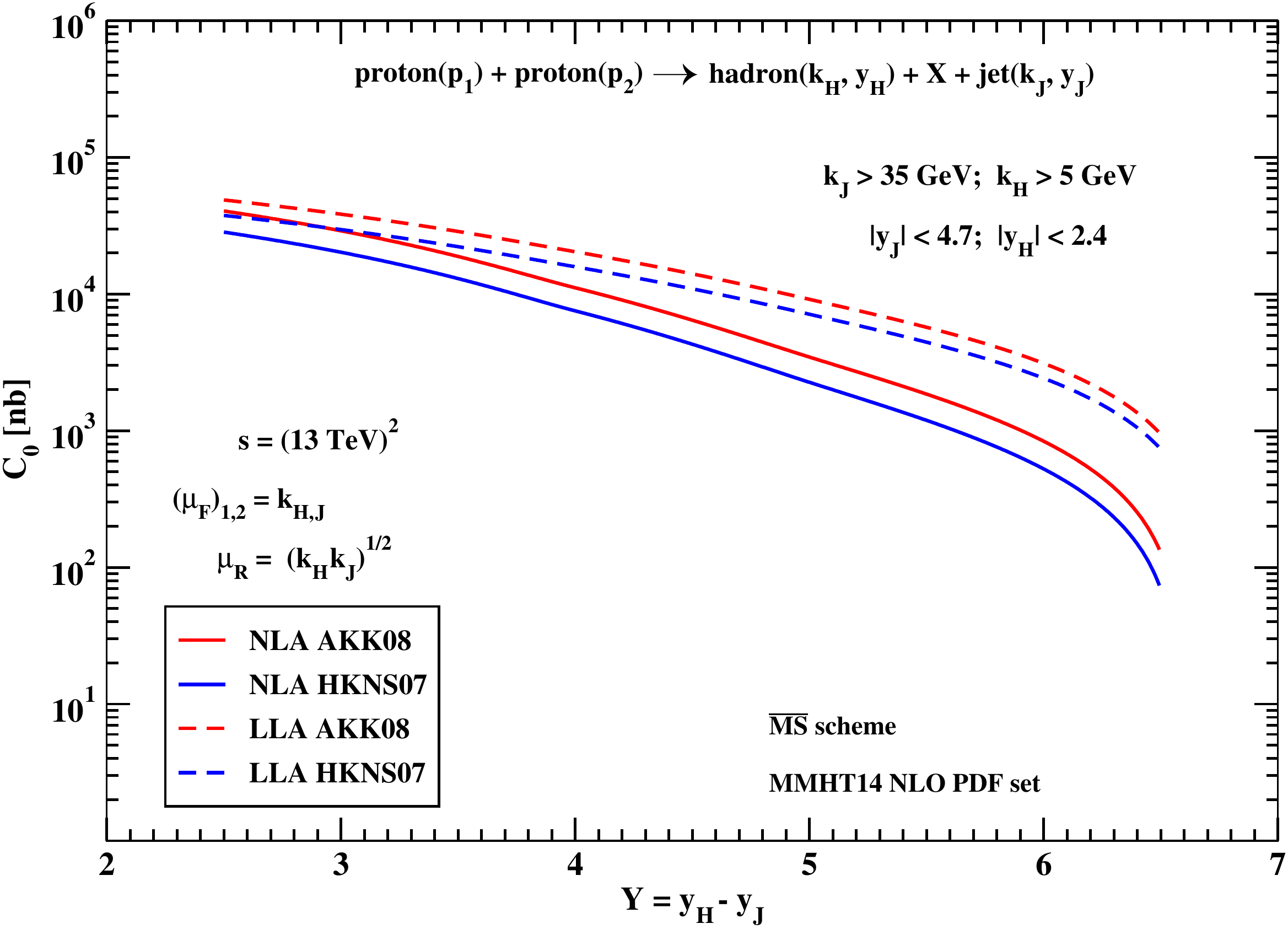}
  \caption{$Y$-dependence of $C_0$ for $\mu_R = \mu_N =
    \sqrt{|\vec k_H||\vec k_J|}$, $(\mu_F)_{1,2} = |\vec k_{H,J}|$, for
    $\sqrt{s}= 7$ TeV (left) and $\sqrt{s} = 13$ TeV (right), and
    $Y \leq 7.1$ ({\it CMS-jet}
    configuration).}	
  \label{fig:C0_MSb_NS_CMS}	
\end{figure}

\begin{figure}[t]
  \centering
  \includegraphics[scale=0.33,clip]{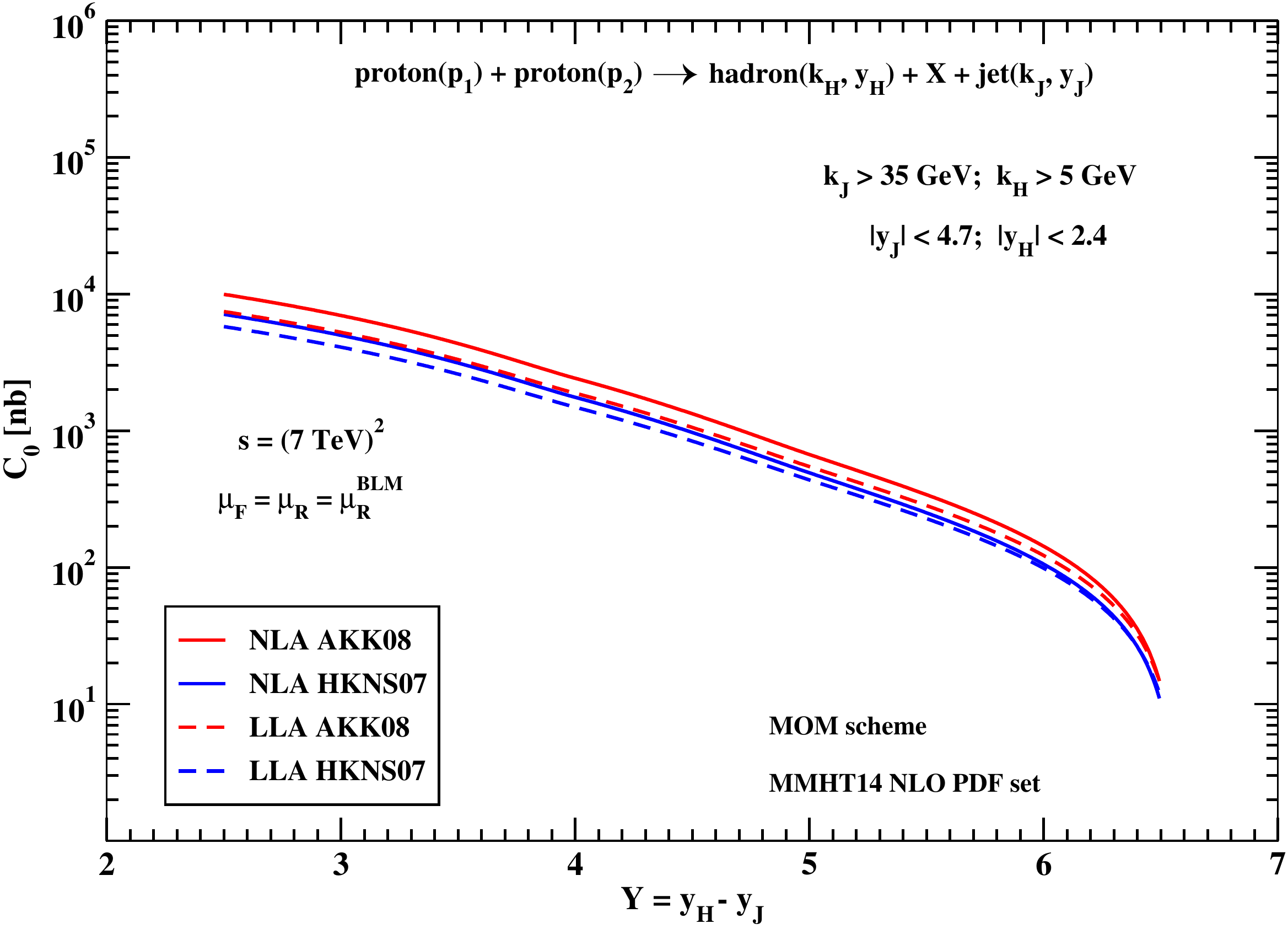}
  \includegraphics[scale=0.33,clip]{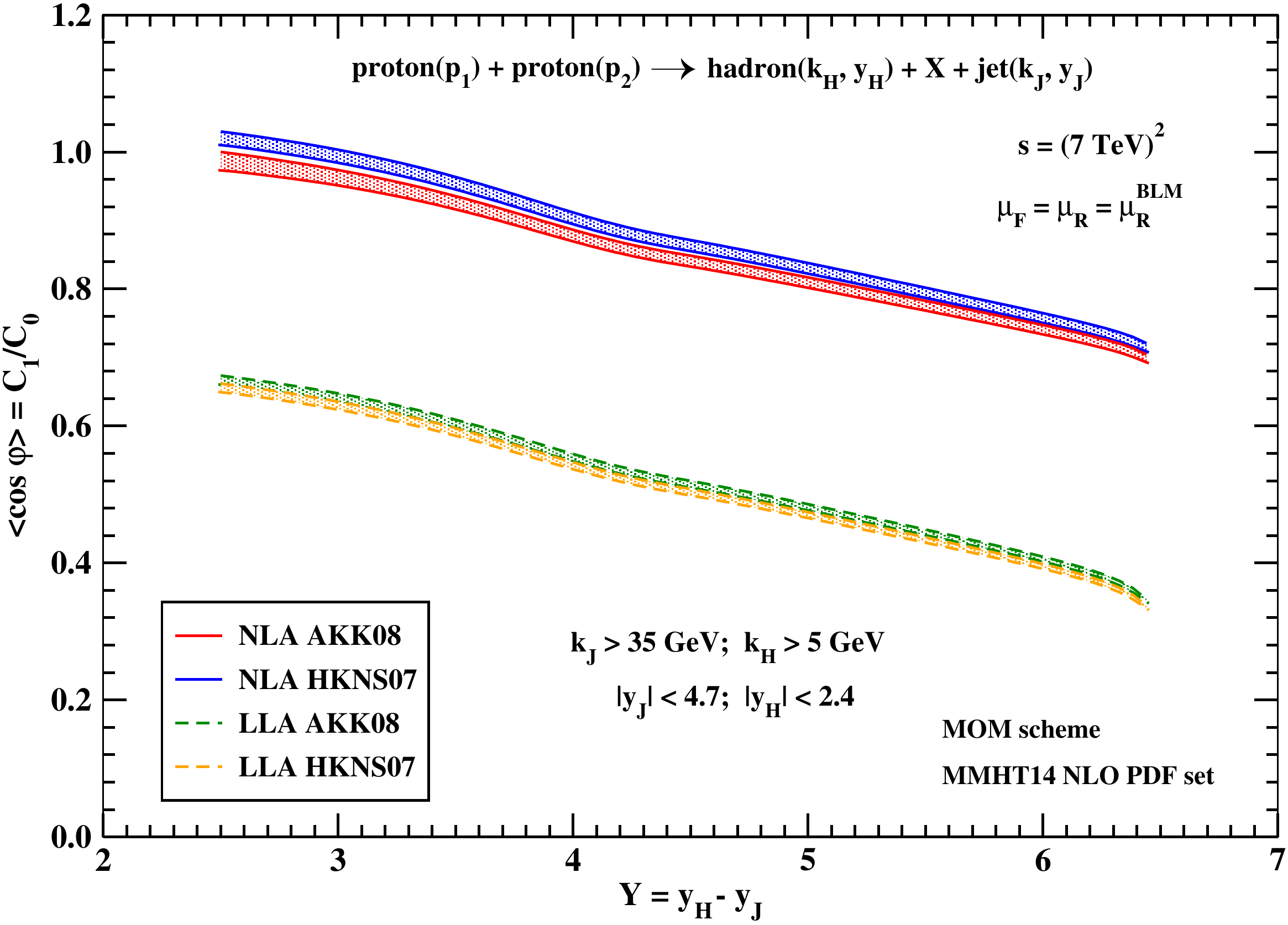}

  \includegraphics[scale=0.33,clip]{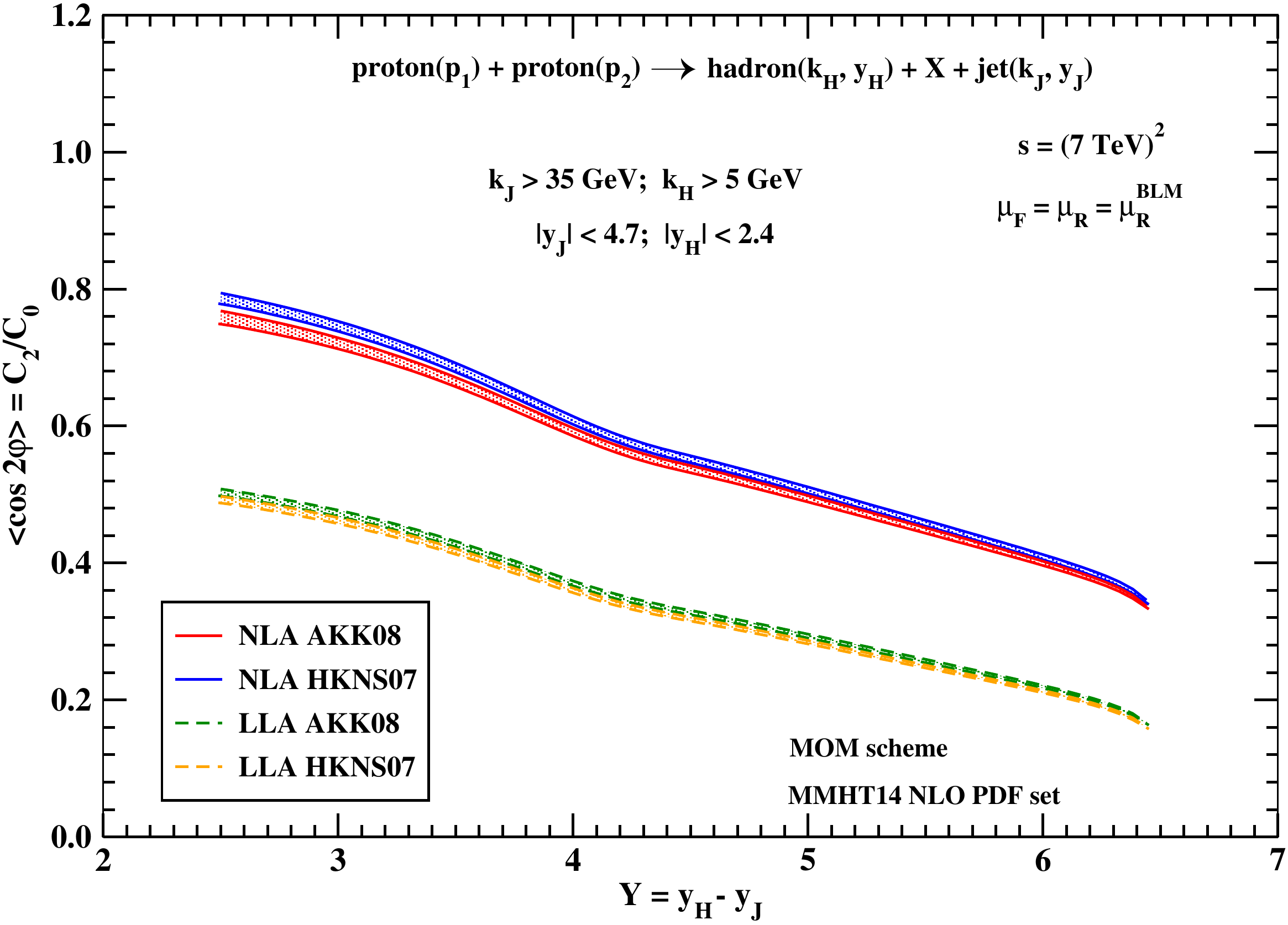}
  \includegraphics[scale=0.33,clip]{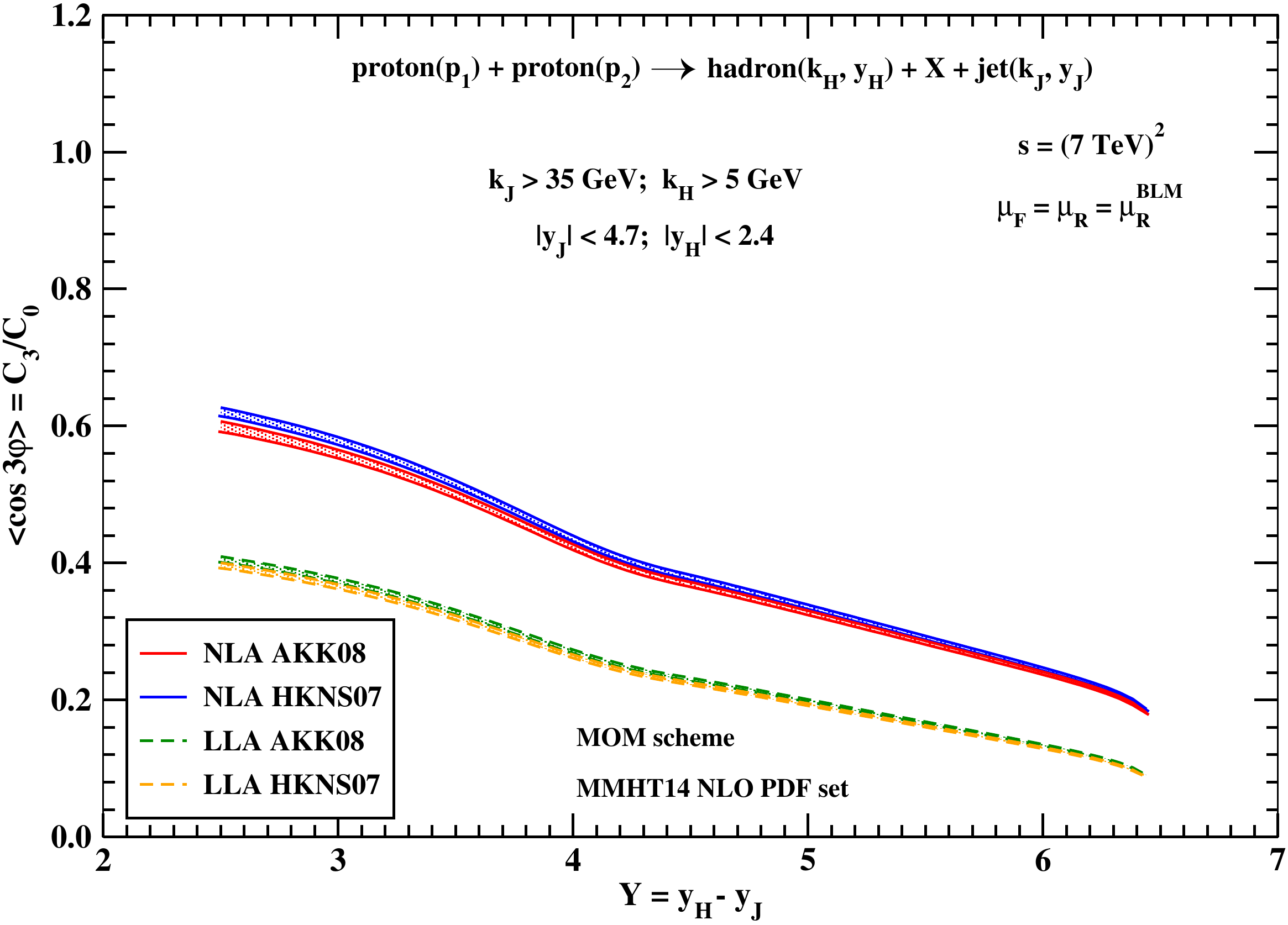}

  \includegraphics[scale=0.33,clip]{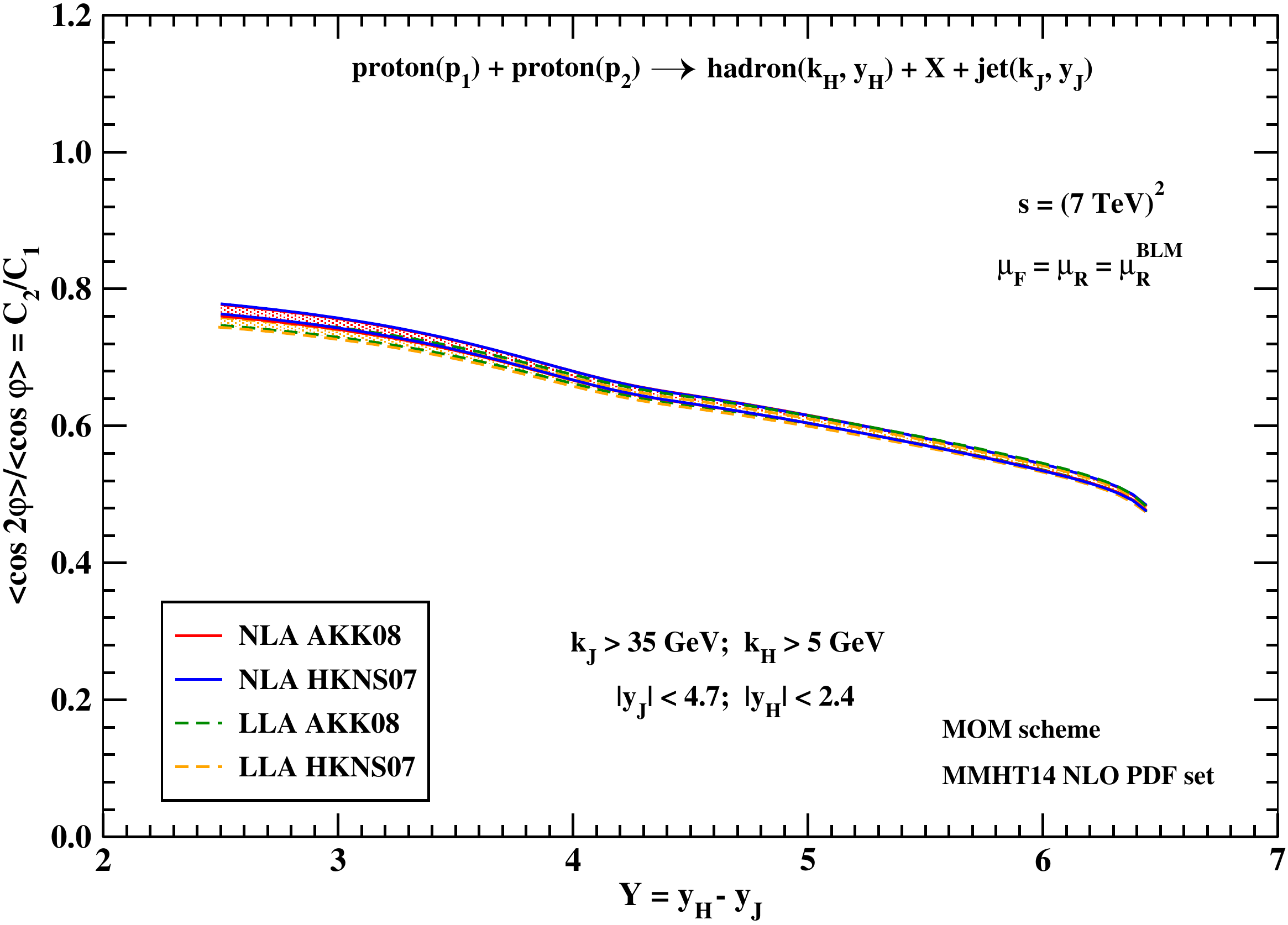}
  \includegraphics[scale=0.33,clip]{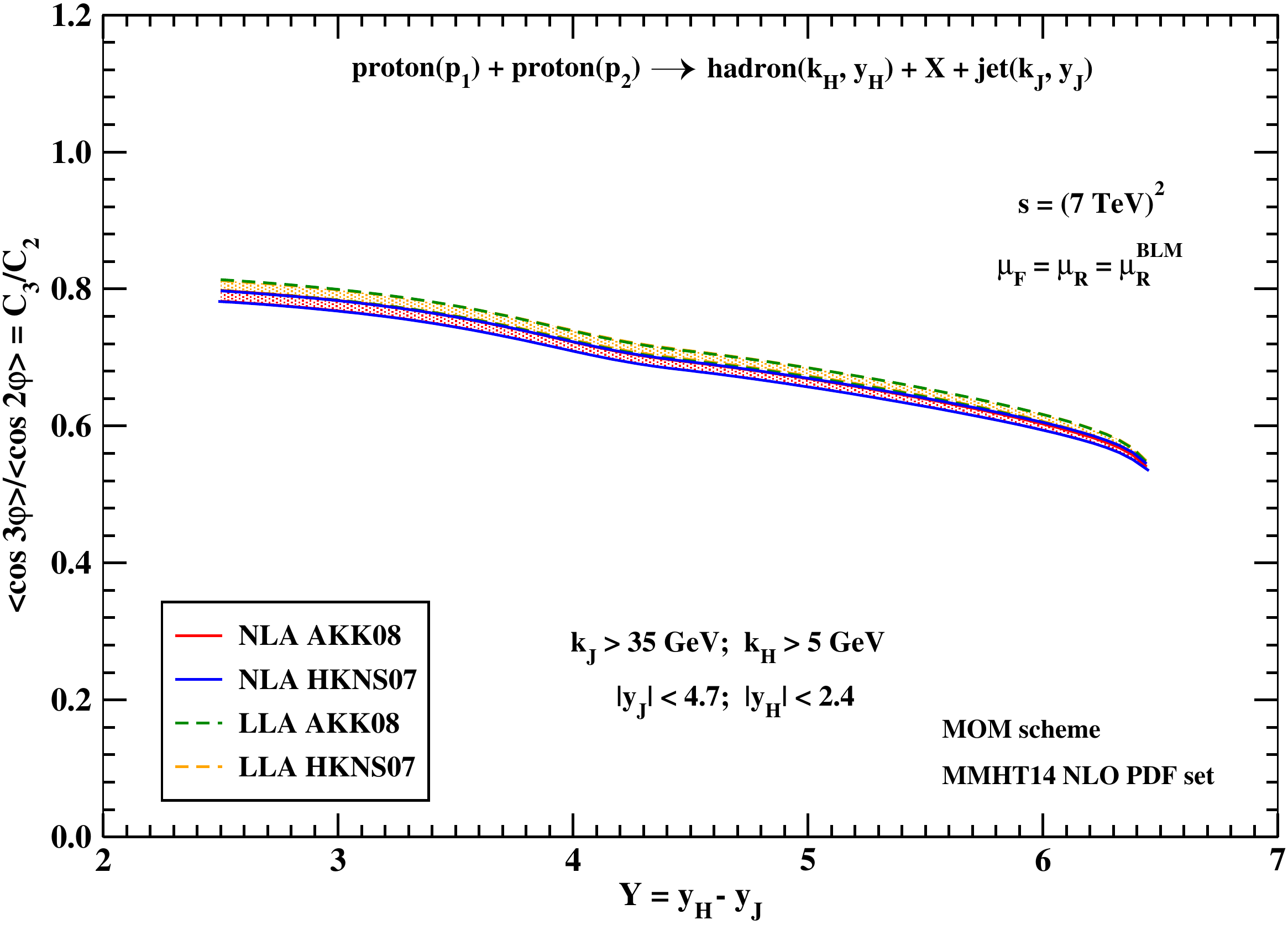}
  \caption{$Y$-dependence of $C_0$ and of several ratios $C_m/C_n$ for 
    $\mu_F = \mu_R^{\rm BLM}$, $\sqrt{s} = 7$ TeV, and $Y \leq 7.1$
    ({\it CMS-jet} configuration).}
  \label{fig:Cn_MOM_BLM_CMS_7}
\end{figure}

\begin{figure}[t]
  \centering
  \includegraphics[scale=0.33,clip]{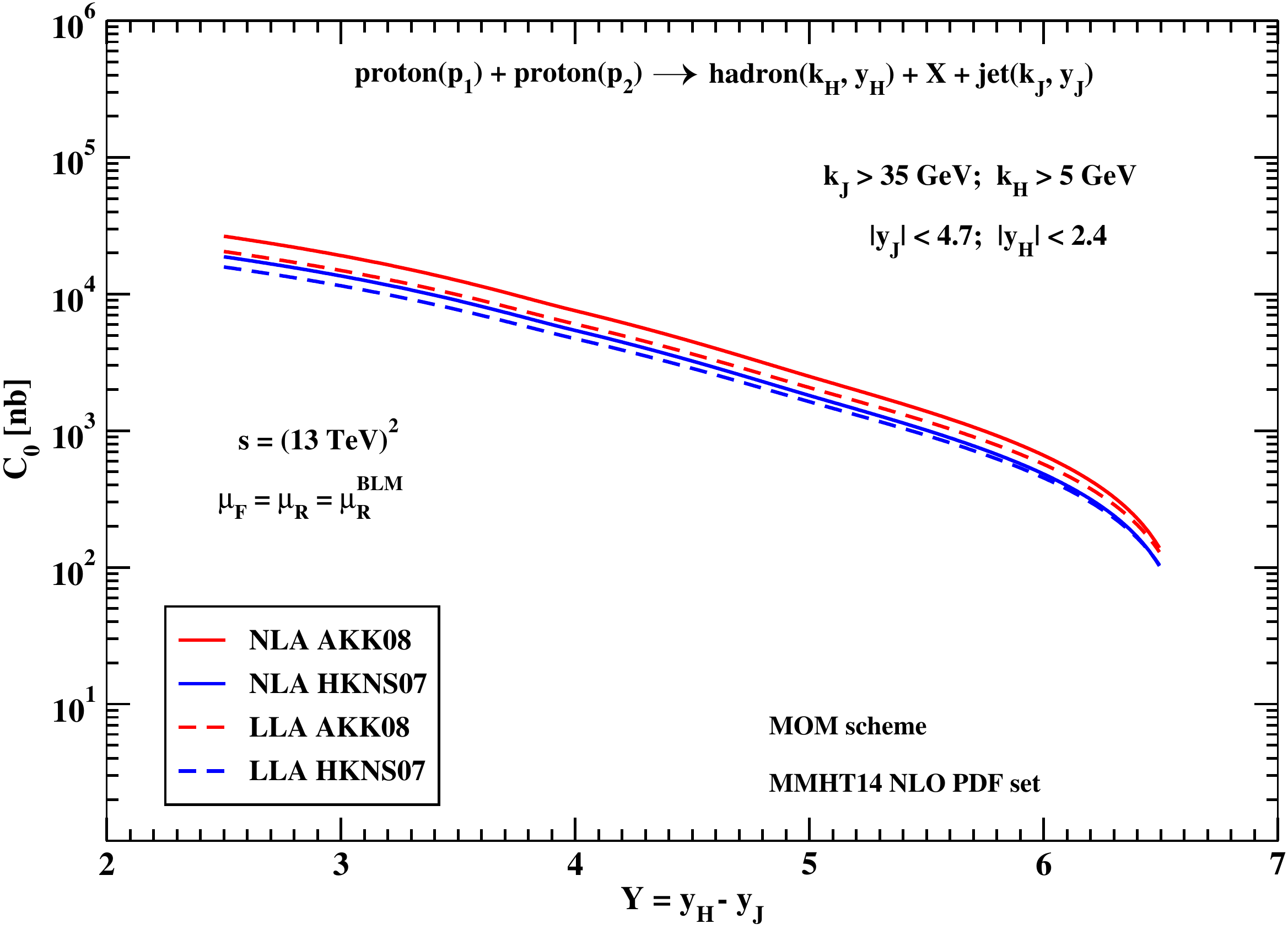}
  \includegraphics[scale=0.33,clip]{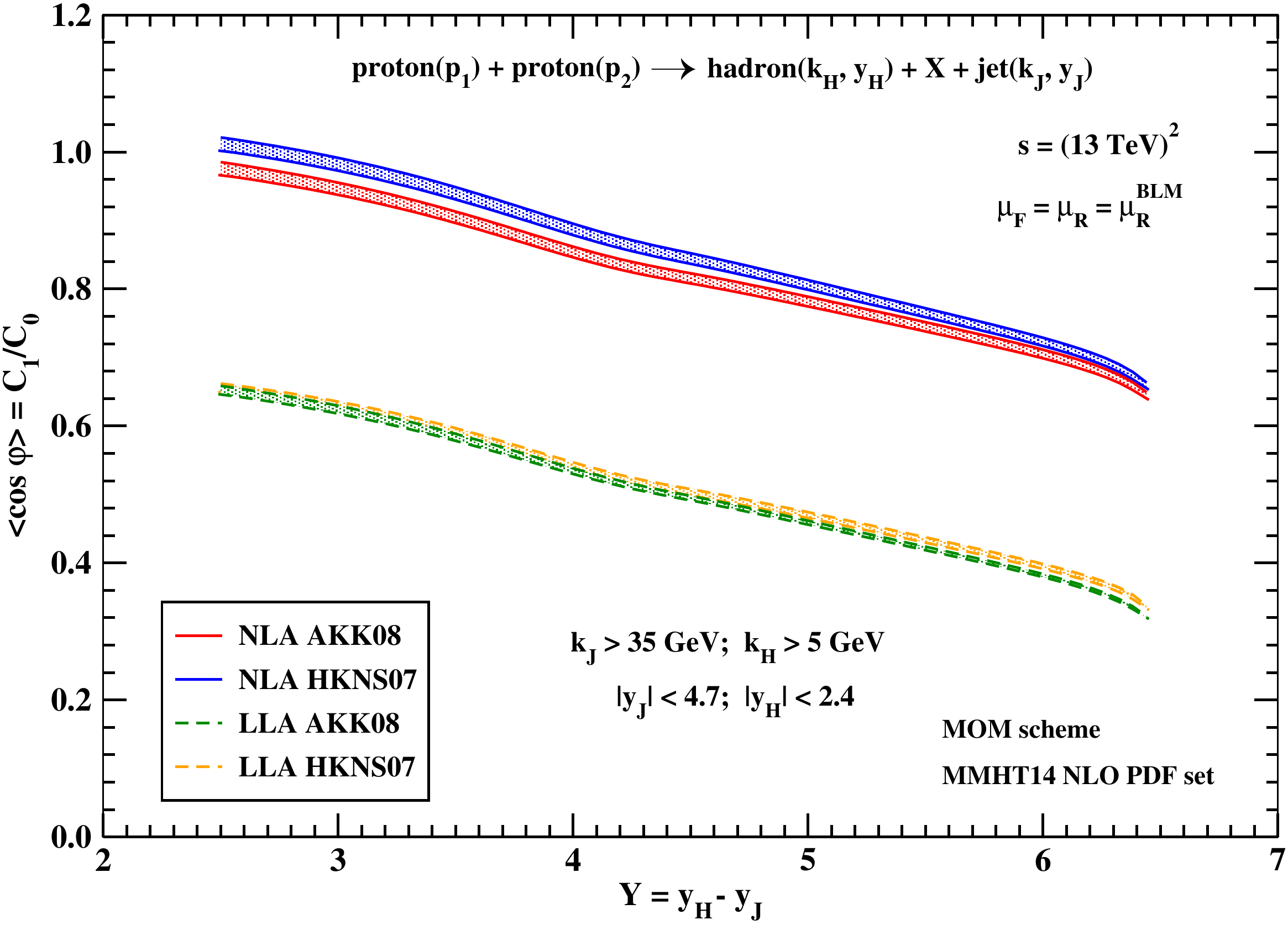}

  \includegraphics[scale=0.33,clip]{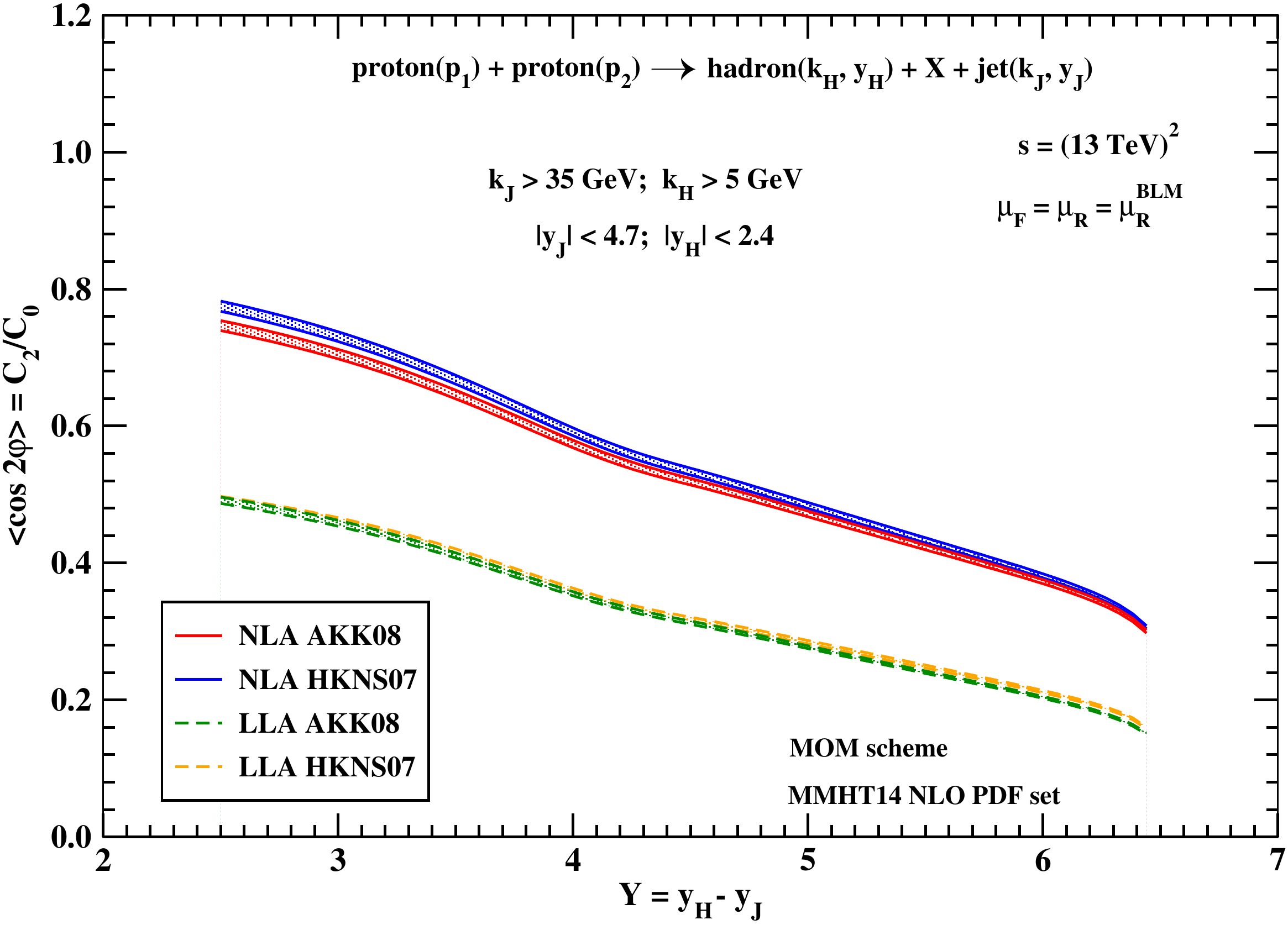}
  \includegraphics[scale=0.33,clip]{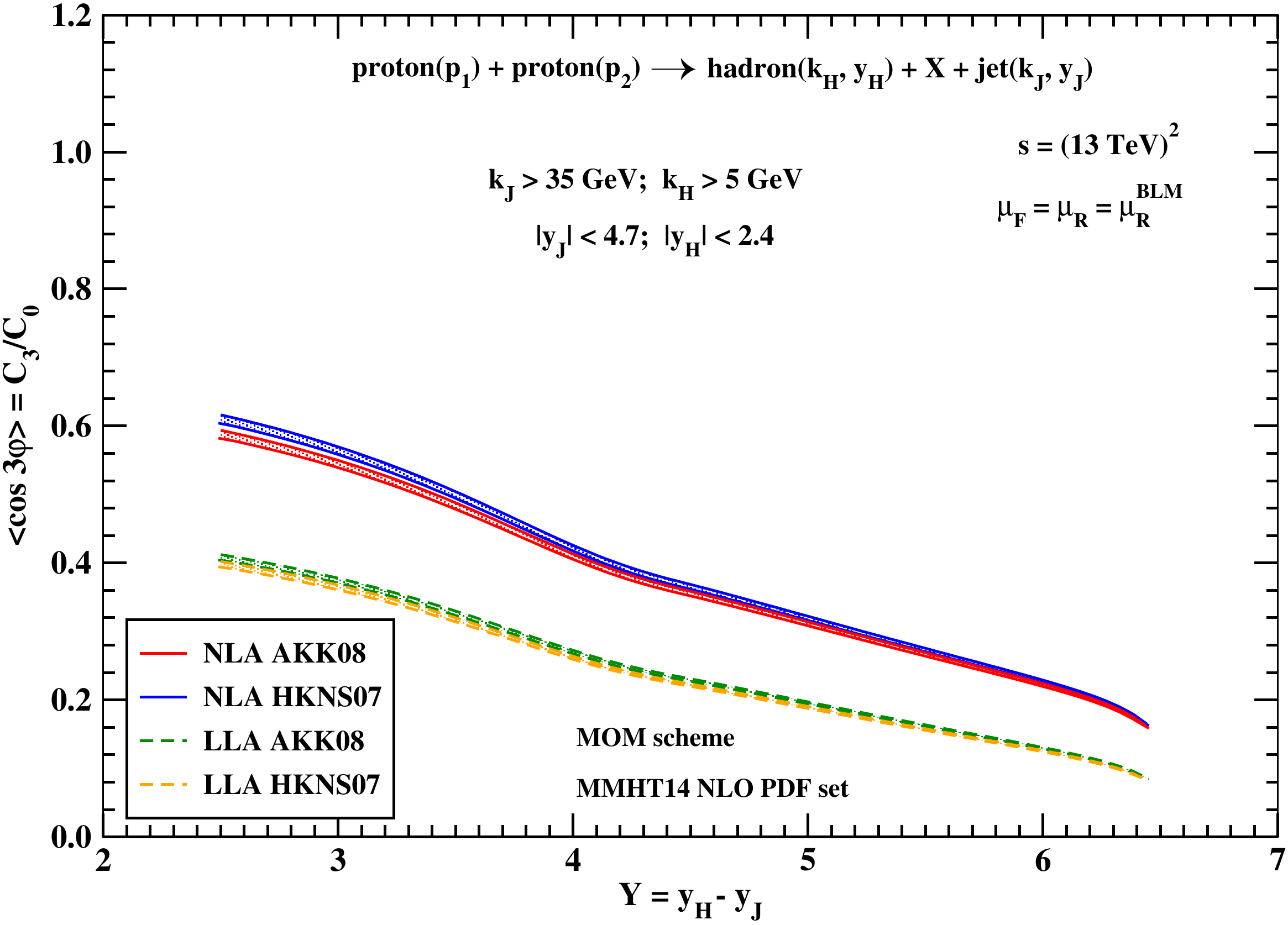}

  \includegraphics[scale=0.33,clip]{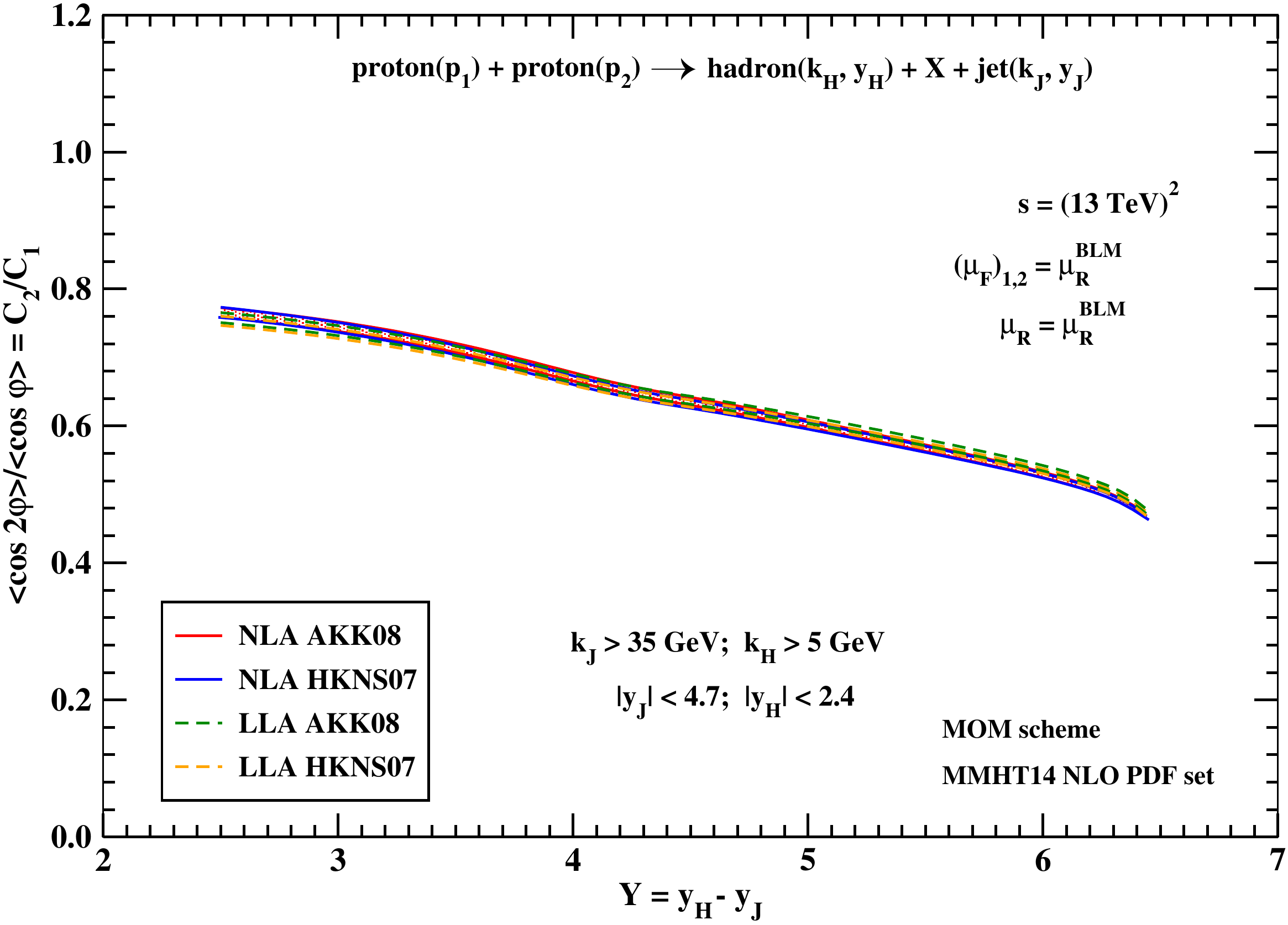}
  \includegraphics[scale=0.33,clip]{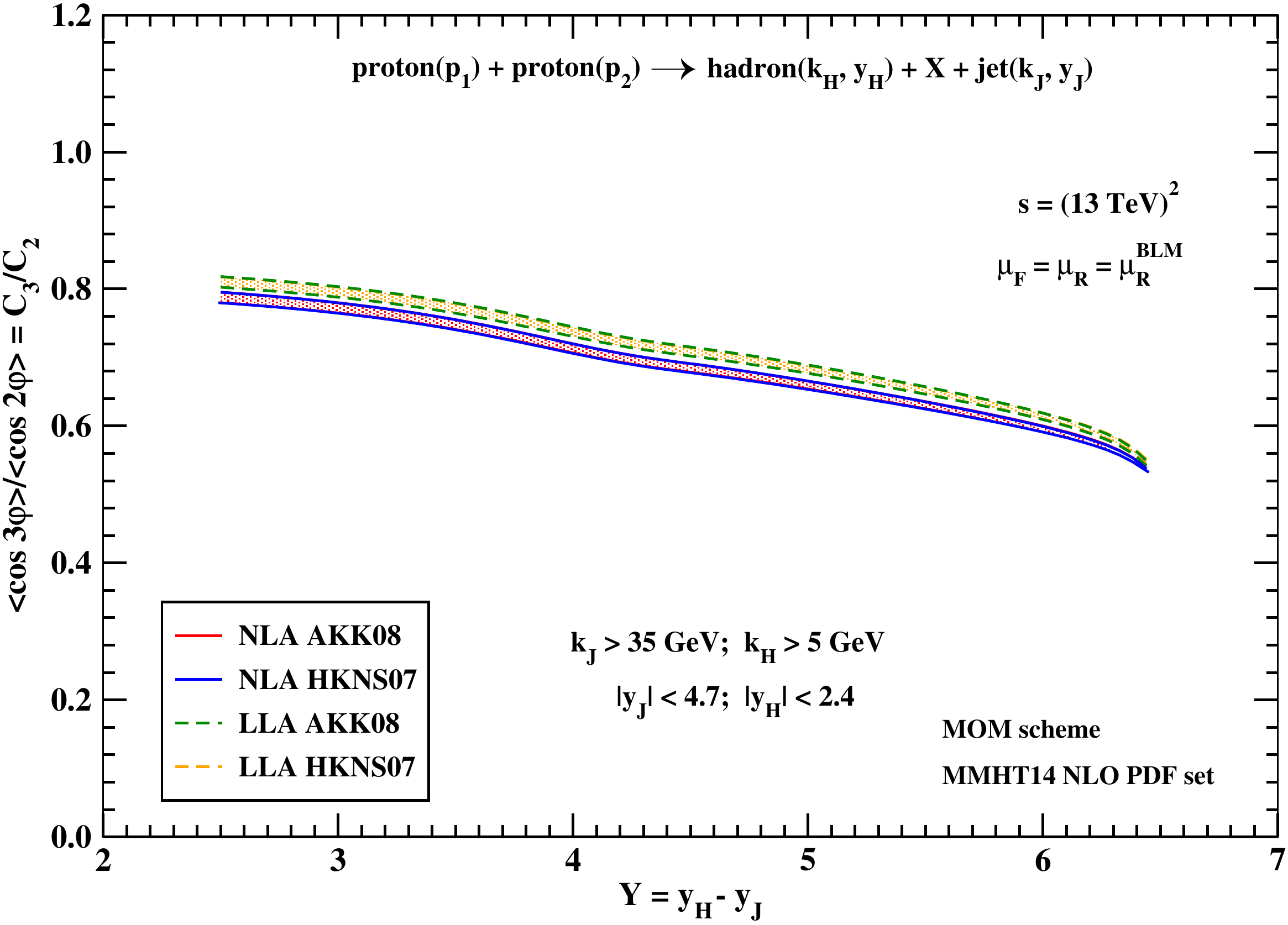}
  \caption{$Y$-dependence of $C_0$ and of several ratios $C_m/C_n$ for 
    $\mu_F = \mu_R^{\rm BLM}$, $\sqrt{s} = 13$ TeV, and $Y \leq 7.1$
    ({\it CMS-jet} configuration).}
  \label{fig:Cn_MOM_BLM_CMS_13}
\end{figure}

\begin{figure}[t]
  \centering
  \includegraphics[scale=0.33,clip]{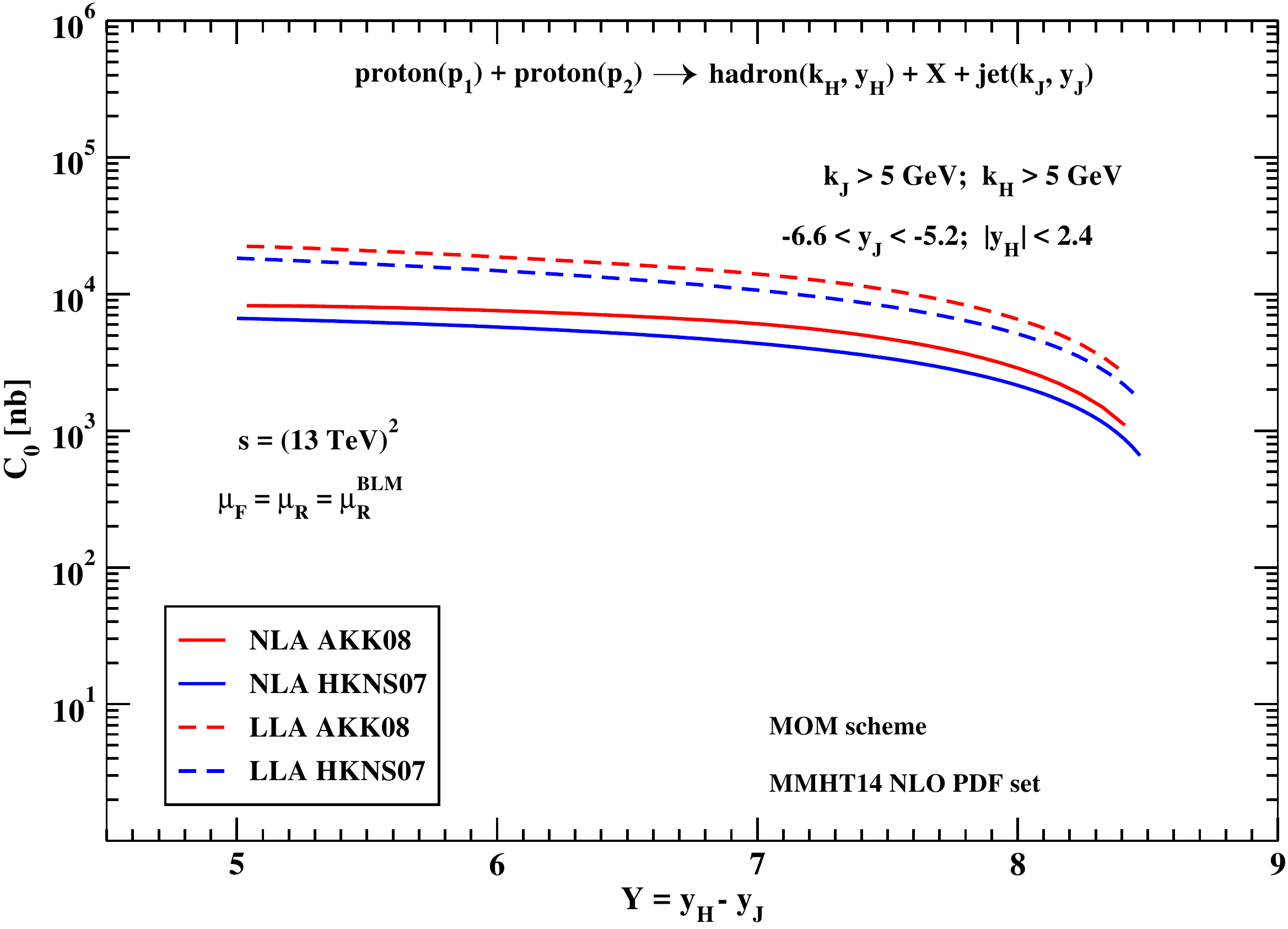}
  \includegraphics[scale=0.33,clip]{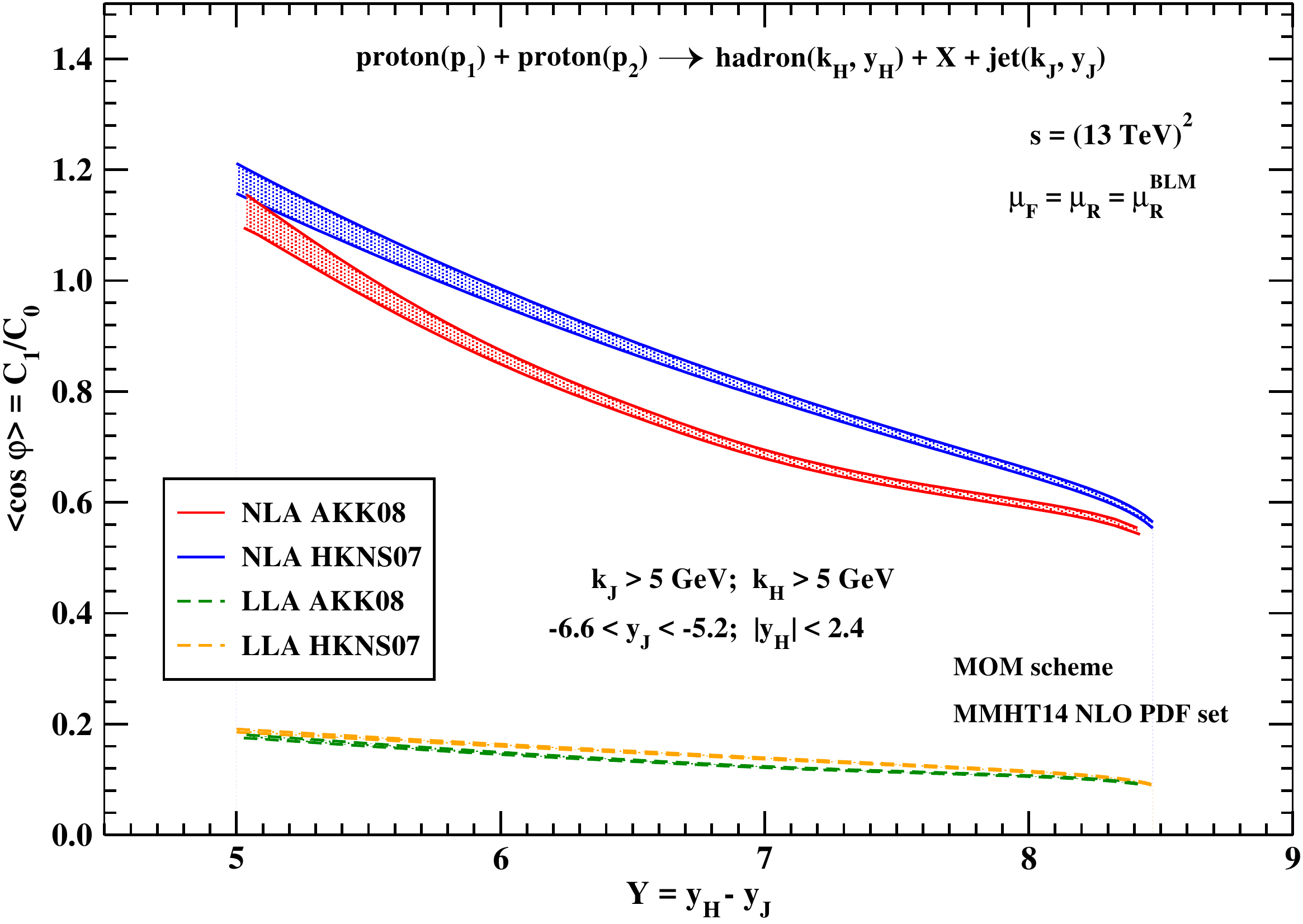}
  
  \includegraphics[scale=0.33,clip]{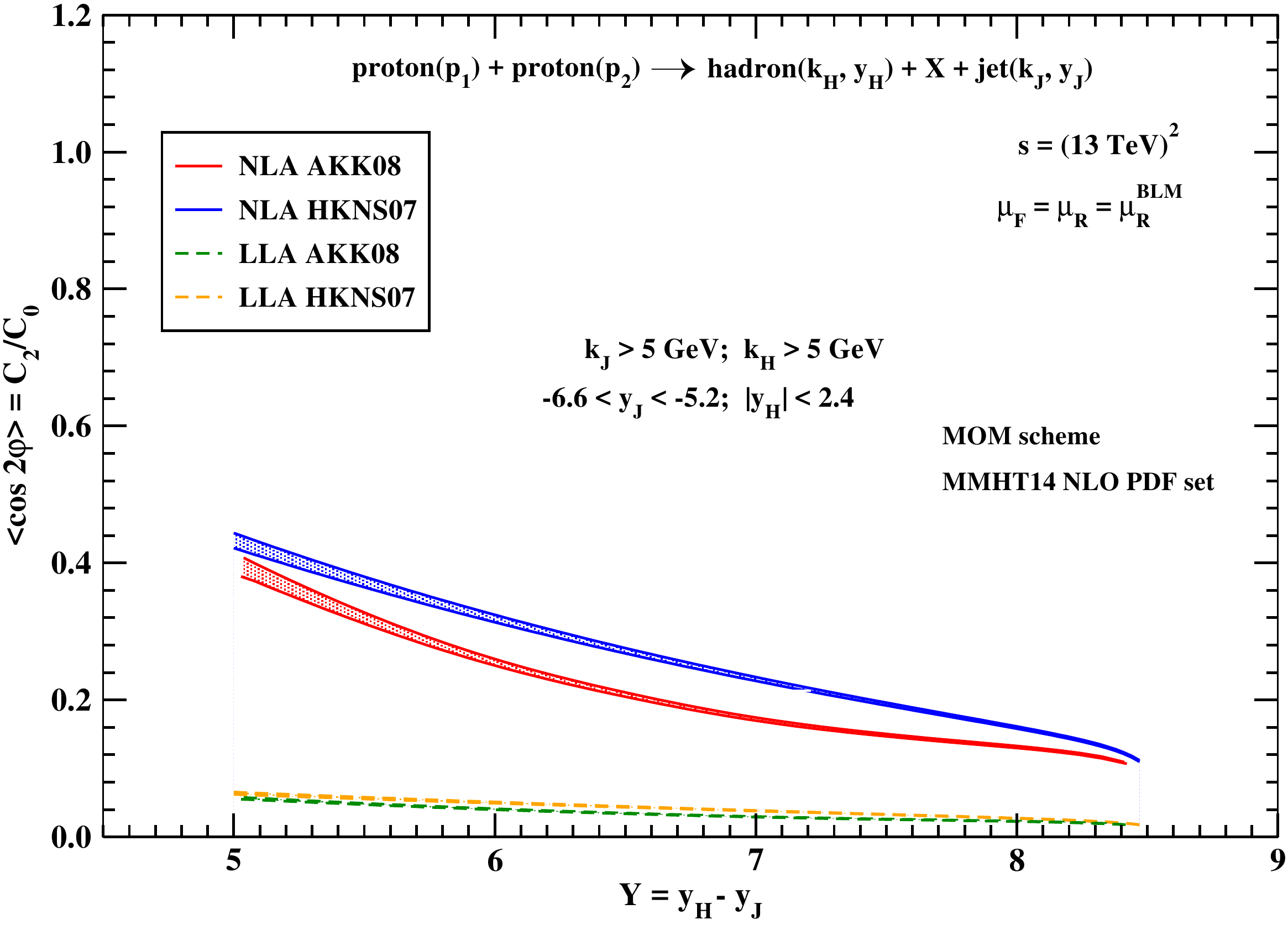}
  \includegraphics[scale=0.33,clip]{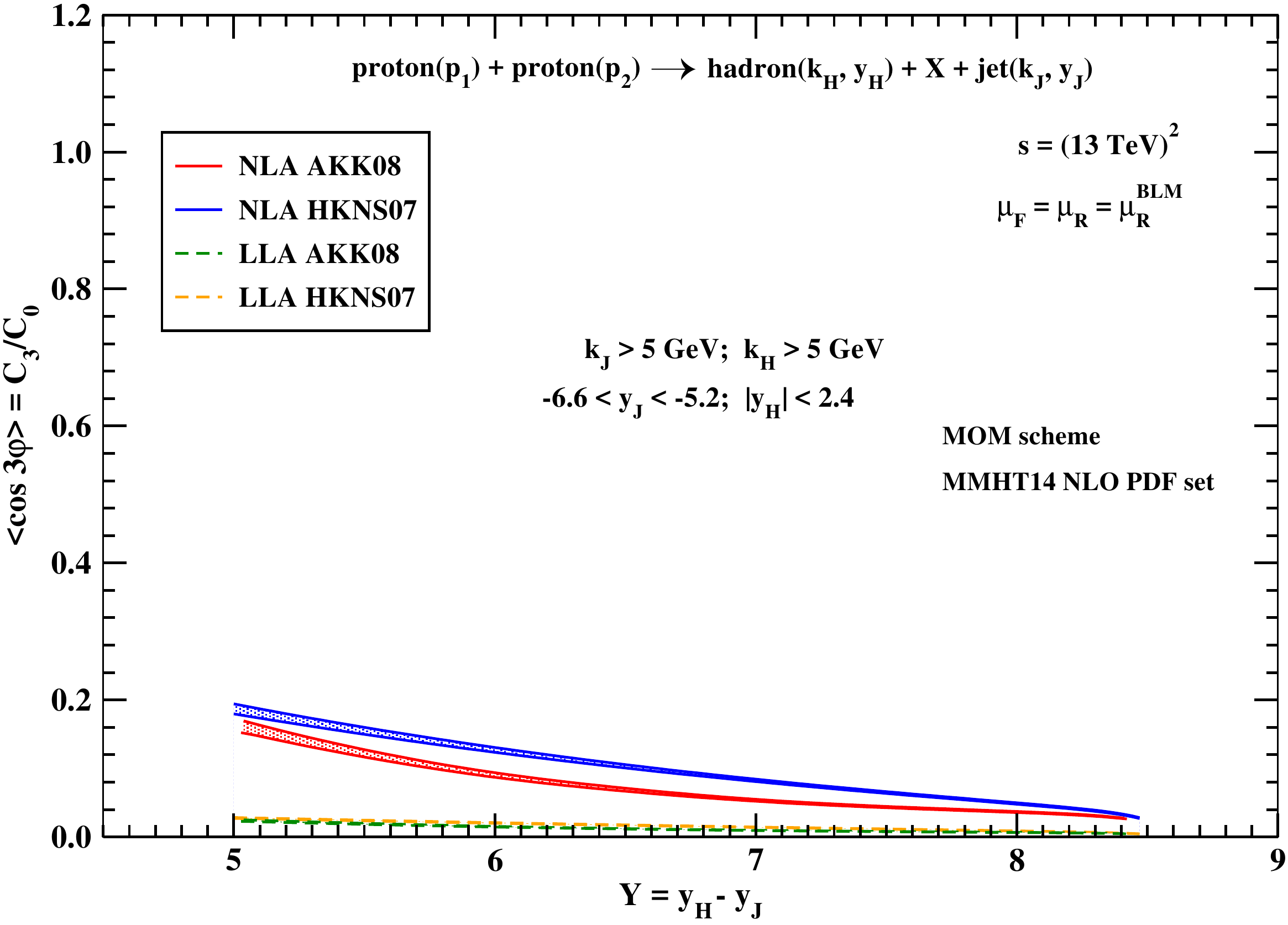}
  
  \includegraphics[scale=0.33,clip]{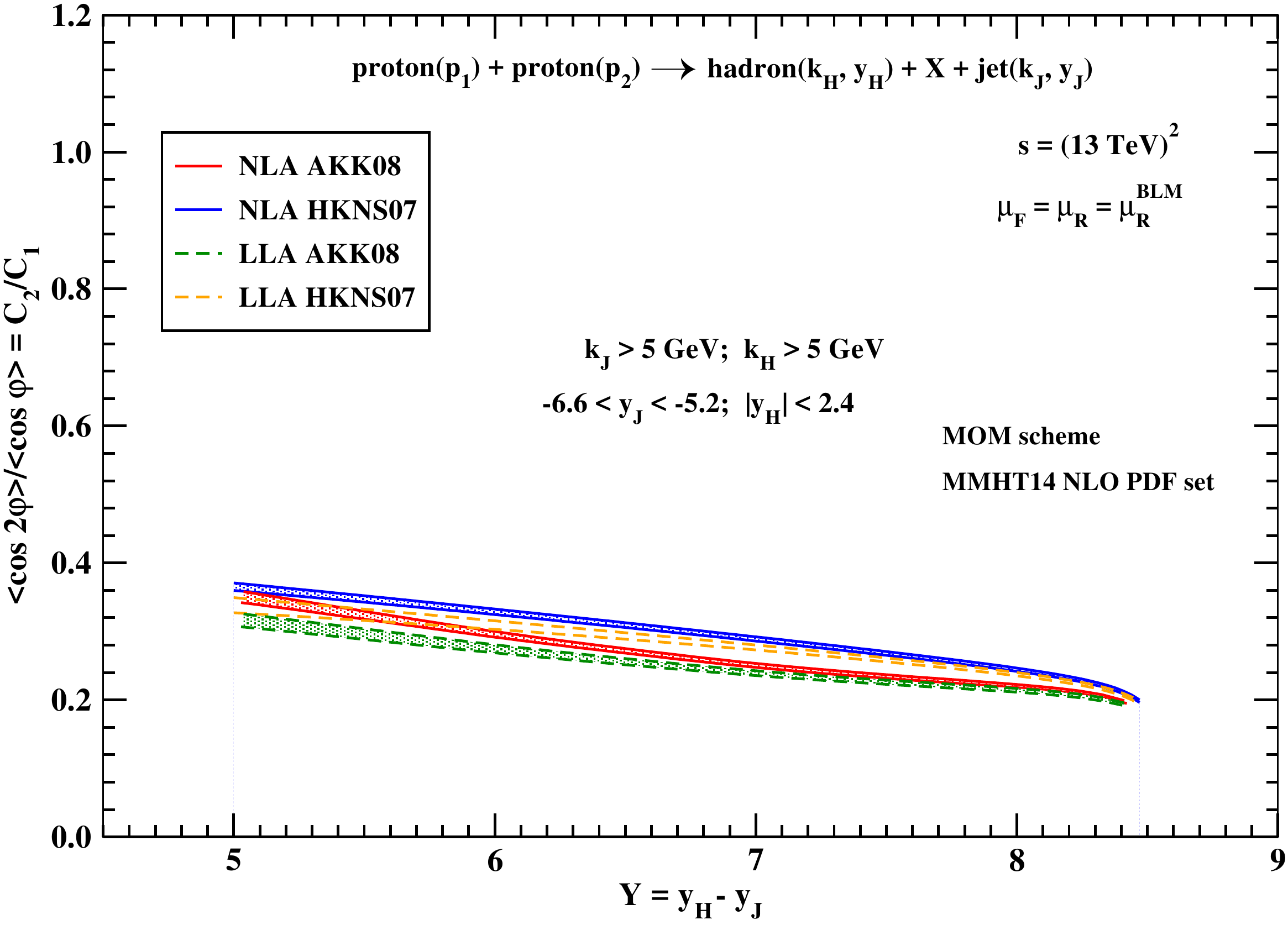}
  \includegraphics[scale=0.33,clip]{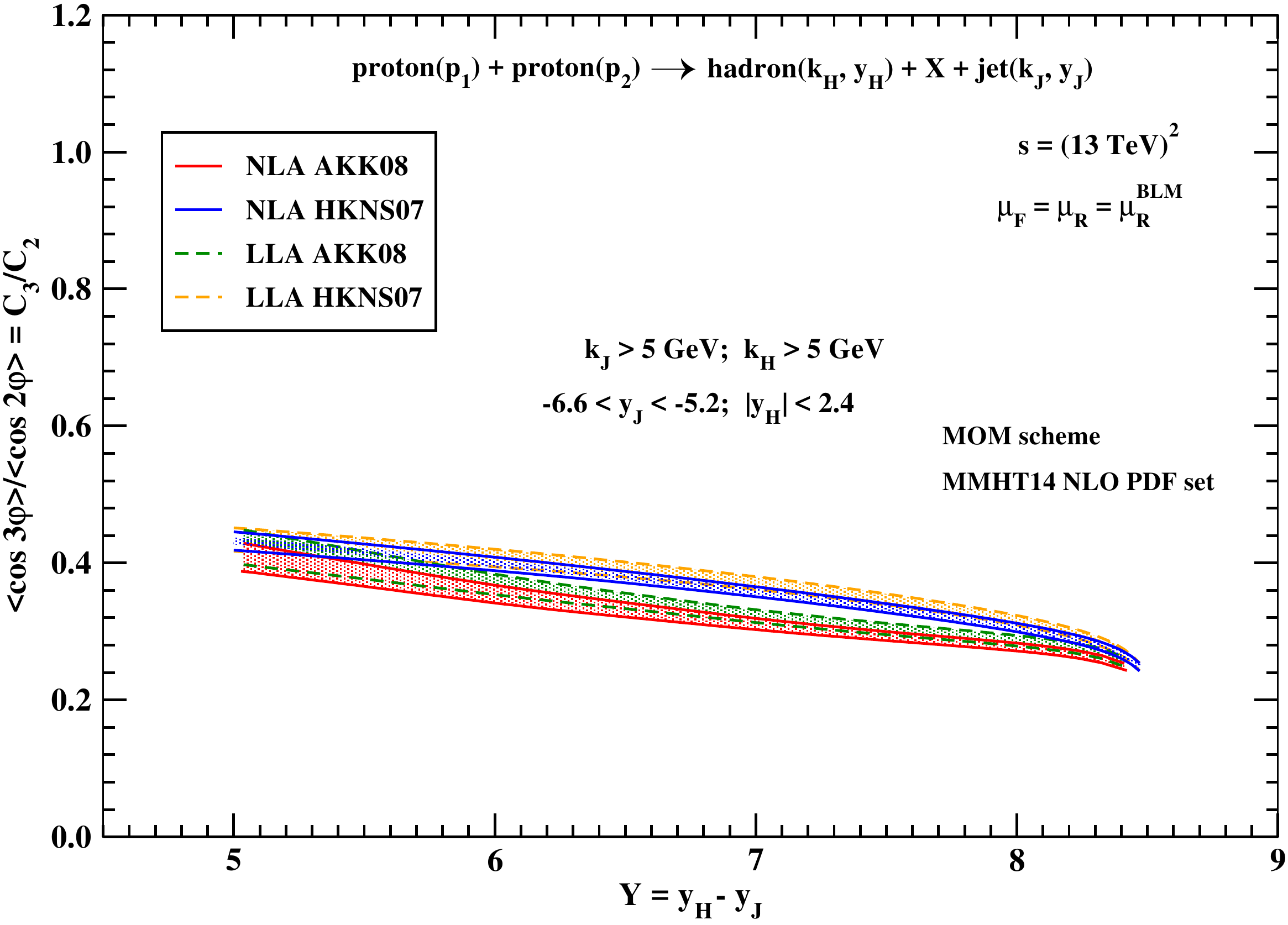}
  \caption{$Y$-dependence of $C_0$ and of several ratios $C_m/C_n$ for 
    $\mu_F = \mu_R^{\rm BLM}$, $\sqrt{s} = 13$ TeV, and $Y \leq 9$
    ({\it CASTOR-jet} configuration).}
  \label{fig:Cn_MOM_BLM_CASTOR_13}
\end{figure}

\begin{figure}[t]
  \centering
  \includegraphics[scale=0.33,clip]{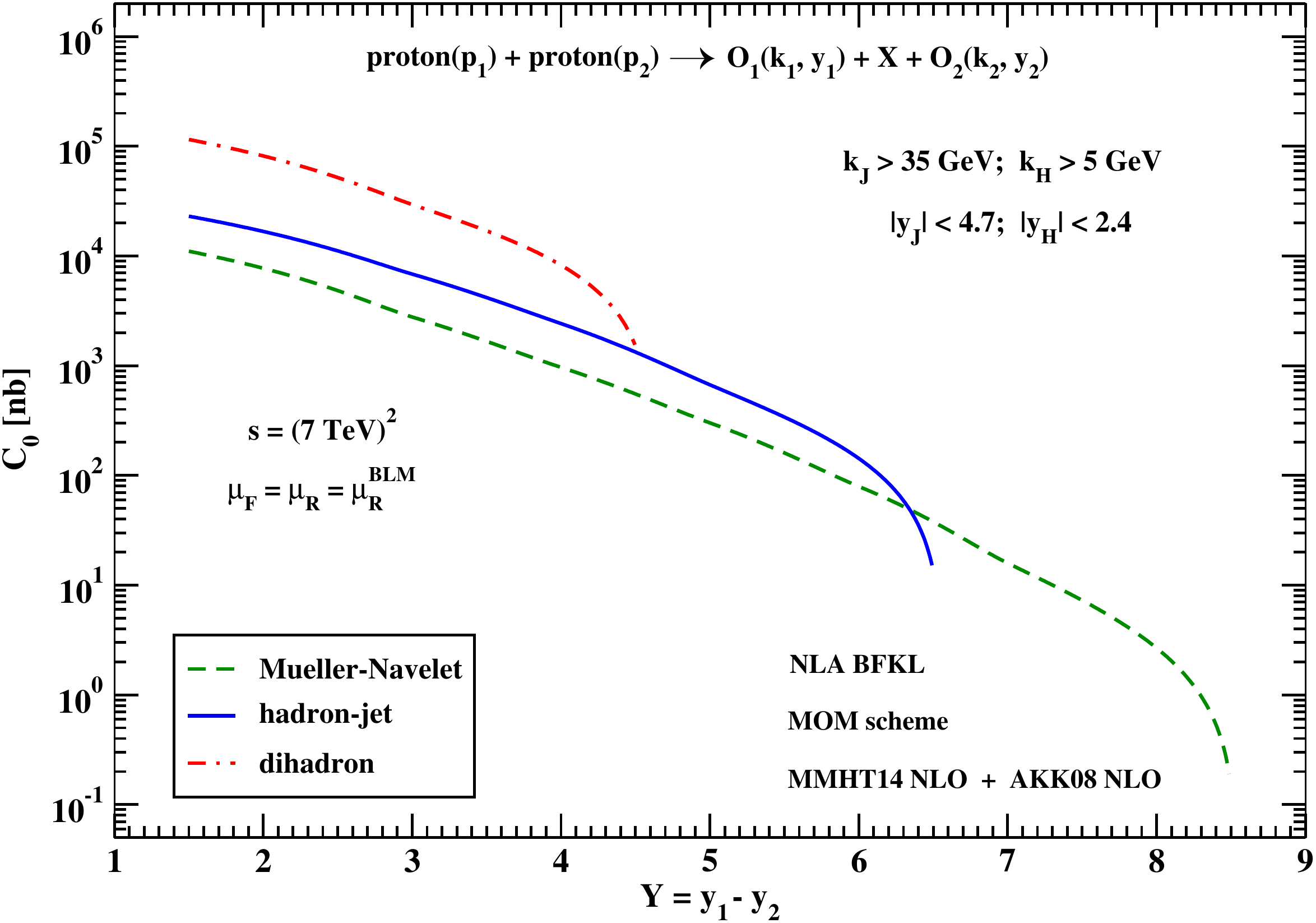}
  \includegraphics[scale=0.33,clip]{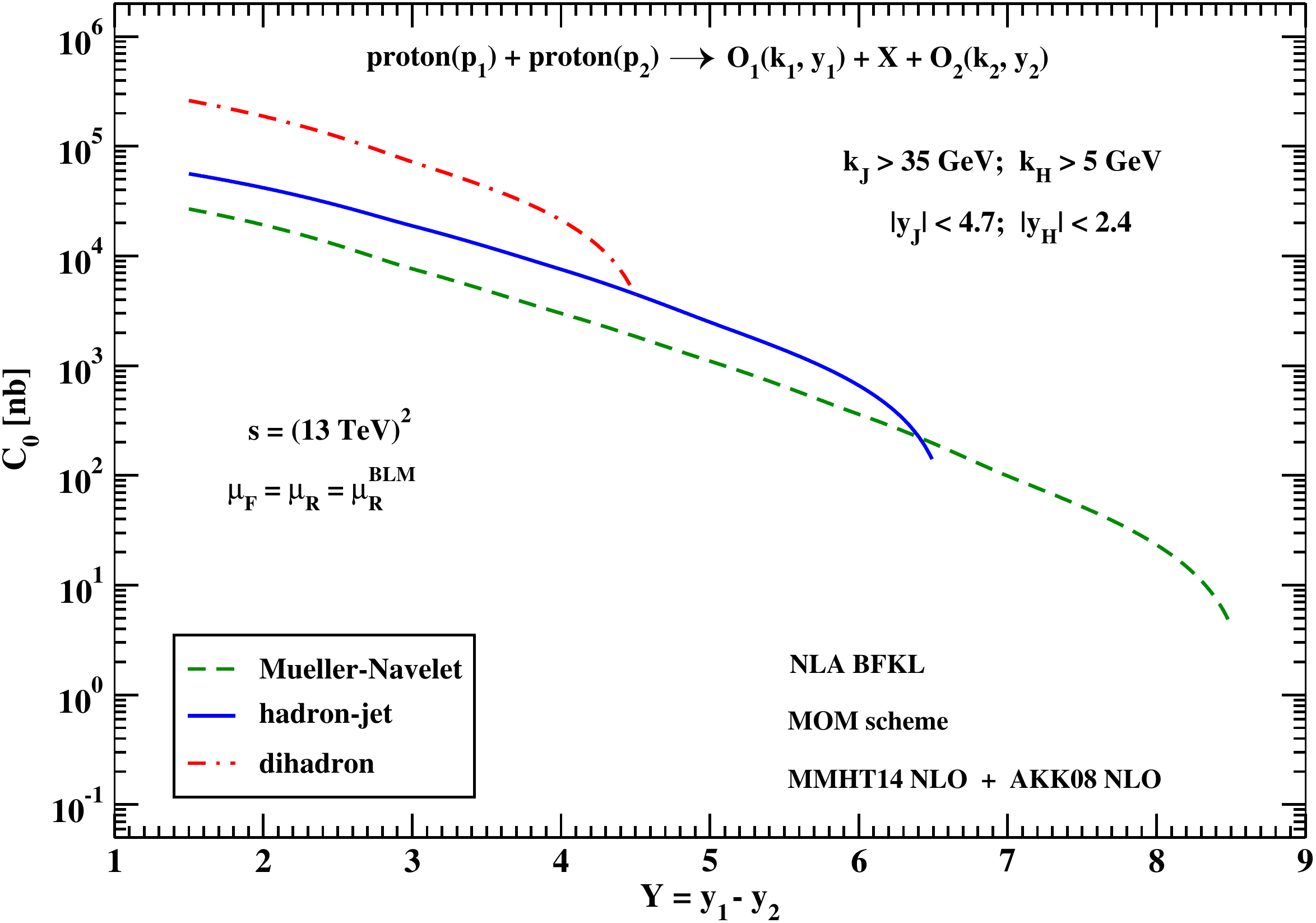}	
  \caption{Comparison of the $\phi$-averaged cross section $C_0$ in different
    NLA BFKL processes: Mueller-Navelet jet, hadron-jet and dihadron production,
    for $\mu_F = \mu_R^{\rm BLM}$, $\sqrt{s} = 7$ and 13 TeV, and $Y \leq 7.1$
    ({\it CMS-jet} configuration).}
  \label{fig:C0_comp_NLA_BLM_CMS}
\end{figure}

\subsection{Integration over the final-state phase space}
\label{phase_space}

In order to match the actual LHC kinematic cuts, we integrate the coefficients
over the phase space for two final-state objects and keep fixed the rapidity
interval, $Y$, between the hadron and the jet: 
\beq
\label{Cn_int}
C_n= 
\int_{y^{\rm min}_H}^{y^{\rm max}_H}dy_H
\int_{y^{\rm min}_J}^{y^{\rm max}_J}dy_J\int_{k^{\rm min}_H}^{k^{\rm max}_H}dk_H
\int_{k^{\rm min}_J}^{{k^{\rm max}_J}}dk_J
\, \delta \left( y_H - y_J - Y \right)
\, {\cal C}_n \left(y_H,y_J,k_H,k_J \right)\, .
\eeq
We consider two distinct ranges for the final-state objects:
\begin{itemize}
\item
  both the hadron and the jet tagged by the CMS detector in their typical
  kinematic configurations, {\it i.e.}:
  $k^{\rm min}_H=5$ GeV, $k^{\rm min}_J=35$ GeV, 
  $y^{\rm max}_H=-y^{\rm min}_H=2.4$, 
  $y^{\rm max}_J=-y^{\rm min}_J=4.7$~\cite{Khachatryan:2016udy}. For the sake of
  brevity, we will refer to this choice as the \textit{\textbf{CMS-jet}}
  configuration; \,
\item
  a hadron always detected inside CMS in the range given above, together with a
  very backward jet tagged by CASTOR. In this peculiar,
  \textit{\textbf{CASTOR-jet}} configuration, the jet lies in the typical
  range of the CASTOR experimental analyses, {\it i.e.}
  $k^{\rm min}_J=5$ GeV, 
  $y^{\rm max}_J=-5.2$, $y^{\rm min}_J=-6.6$ ~\cite{CMS:2016ndp},
\end{itemize}

The value of $k_H^{\rm max}$ is constrained by the lower cutoff of the adopted
FF parametrizations (see below) and is always fixed at 21.5~GeV.
The value of $k_J^{\rm max}$ is instead constrained by the requirement that
$x_J \le 1$ which implies $k_J^{\rm max} \simeq 60$ GeV for $\sqrt{s} = 7$~TeV
and $|y_J| < 4.7$ ({\it CMS-jet)} and $k_J^{\rm max} \simeq 17.68$~GeV for
$\sqrt{s} = 13$~TeV ({\it CASTOR-jet)}.

The rapidity interval, $Y$, is taken to be positive: $0 < Y \leq y^{\rm max}_H
- y^{\rm min}_J$. Two center-of-mass energies, $\sqrt s = 7$ and 13 TeV, are
taken into account in the {\it CMS-jet} configuration, while we give
predictions for $\sqrt s = 13$ TeV in the {\it CASTOR-jet} case.

In our calculations we use the MMHT 2014 NLO PDF set~\cite{Harland-Lang:2014zoa}
with two different NLO parametrizations for hadron FFs:
AKK~2008~\cite{Albino:2008fy} and HKNS~2007~\cite{Hirai:2007cx}
(see Section~\ref{numerics} for a related discussion). In the results
presented below, we sum over the production of forward charged light 
hadrons: $\pi^{\pm}, K^{\pm}, p,\bar p$.

\subsection{Scale optimization}
\label{scale_optimization}

To fix the renormalization scale $\mu_R$, which can be arbitrarily chosen
within the NLA, we adopt the
BLM~\cite{Brodsky:1996sgBrodsky:1997sdBrodsky:1998knBrodsky:2002ka}
approach, which has become a quite common choice for semihard processes.
We first perform a finite renormalization from the $\overline{\rm MS}$ to
the physical MOM scheme, whose definition is related to the 3-gluon vertex
being a key ingredient of the BFKL approach and get
\beq{}
\alpha_s^{\overline{\rm MS}}=\alpha_s^{\rm MOM}\left(1+\frac{\alpha_s^{\rm MOM}}{\pi}T
\right)\;,
\eeq
with $T=T^{\beta}+T^{\rm conf}$,
\beq{}
T^{\beta}=-\frac{\beta_0}{2}\left( 1+\frac{2}{3}I \right)\, ,
\eeq
\[ 
T^{\rm conf}= \frac{3}{8}\left[ \frac{17}{2}I +\frac{3}{2}\left(I-1\right)\xi
+\left( 1-\frac{1}{3}I\right)\xi^2-\frac{1}{6}\xi^3 \right] \;,
\]
where $I=-2\int_0^1dx\frac{\ln\left(x\right)}{x^2-x+1}\simeq2.3439$ and $\xi$
is the gauge parameter of the MOM scheme, fixed at zero in the following.
Then, the ``optimal'' BLM scale $\mu_R^{\rm BLM}$ is the value of $\mu_R$
that makes the $\beta_0$-dependent part in the expression for the observable
of interest vanish.  
In Ref.~\cite{Caporale:2015uva} some of us showed that terms proportional 
to the QCD $\beta_0$-function are present not only in the NLA BFKL kernel, 
but also in the expressions for the NLA impact factor. This leads 
to a non-universality of the BLM scale and to its dependence on the 
energy of the process. 

Finally, the condition for the BLM scale setting was found to be 
\[
C^{\beta}_n
\propto \!\!
\int_{y^{\rm min}_H}^{y^{\rm max}_H}dy_H
\int_{y^{\rm min}_J}^{y^{\rm max}_J}dy_J\int_{k^{\rm min}_H}^{k^{\rm max}_H}dk_H
\int_{k^{\rm min}_J}^{k^{\rm max}_J}dk_J
\!\! 
\int\limits^{\infty}_{-\infty} \!\!d\nu\,e^{Y \bar \alpha^{\rm MOM}_s(\mu^{\rm BLM}_R)\chi(n,\nu)}
c_H(n,\nu)[c_J(n,\nu)]^*
\]
\[
\left[\frac{5}{3}
+\ln \frac{(\mu^{\rm BLM}_R)^2}{|\vec k_H|
|\vec k_J|} +f(\nu)-2\left( 1+\frac{2}{3}I \right)
\right.
\]
\beq{}
\label{beta0}
\left.
+\bar \alpha^{\rm MOM}_s(\mu^{\rm BLM}_R) Y \: \frac{\chi(n,\nu)}{2}
\left(-\frac{\chi(n,\nu)}{2}+\frac{5}{3}+\ln \frac{(\mu^{\rm BLM}_R)^2}{|\vec k_H|
|\vec k_J|}
+f(\nu)-2\left( 1+\frac{2}{3}I \right)\right)\right]=0 \, .
\eeq
The term in the r.h.s. of~Eq.~(\ref{beta0}) proportional to $\alpha^{\rm MOM}_s$ originates from the NLA part of the kernel, while the remaining ones come from the NLA corrections to the hadron/jet vertices.

In order to find the values of the BLM scales, we introduce the ratios of the 
BLM to the ``natural'' scale suggested by the kinematic of the process, 
$\mu_N=\sqrt{|\vec k_H||\vec k_J|}$, so that $m_R=\mu_R^{\rm BLM}/\mu_N$, 
and look for the values of $m_R$ which solve Eq.~(\ref{beta0}). 

We finally plug these scales into our expression for the integrated
coefficients in the BLM scheme (for the derivation see
Ref.~\cite{Caporale:2015uva}):
\beq
\label{Cn_int_blm}
C_n =
\int_{y^{\rm min}_H}^{y^{\rm max}_H}dy_H
\int_{y^{\rm min}_J}^{y^{\rm max}_J}dy_J\int_{k^{\rm min}_H}^{k^{\rm max}_H}dk_H
\int_{k^{\rm min}_J}^{k^{\rm max}_J}dk_J
\,
\int\limits^{\infty}_{-\infty} d\nu 
\eeq
\beq \nonumber
\frac{e^Y}{s}\,
 e^{Y \bar \alpha^{\rm MOM}_s(\mu^{\rm BLM}_R)\left[\chi(n,\nu)
+\bar \alpha^{\rm MOM}_s(\mu^{\rm BLM}_R)\left(\bar \chi(n,\nu) +\frac{T^{\rm conf}}
{3}\chi(n,\nu)\right)\right]}
\left(\alpha^{\rm MOM}_s (\mu^{\rm BLM}_R)\right)^2 
\eeq
\[
\times c_H(n,\nu)[c_J(n,\nu)]^*
\left\{1+\bar \alpha^{\rm MOM}_s(\mu^{\rm BLM}_R)\left[\frac{\bar c^{(1)}_H(n,\nu)}
  {c_H(n,\nu)}+\left[\frac{\bar c^{(1)}_J(n,\nu)}{c_J(n,\nu)}\right]^*
  +\frac{2T^{\rm conf}}{3} \right] \right\} \, .
\]
The coefficient $C_0$ gives the $\phi$-averaged cross section, while the ratios
$R_{n0} \equiv C_n/C_0 = \langle\cos(n\phi)\rangle$ determine the values of the
mean cosines, or azimuthal correlations, of the produced hadron and jet.
In Eq.~(\ref{Cn_int_blm}), $\bar \chi(n,\nu)$ is the eigenvalue of NLA BFKL
kernel~\cite{Kotikov:2000pm} and its expression is given, {\it e.g.} in
Eq.~(23) of Ref.~\cite{Caporale:2012ih}, whereas $\bar c^{(1)}_{H,J}$ are the
NLA parts of the hadron/jet vertices (see Ref.~\cite{Caporale:2015uva}).

We set the factorization scale $\mu_F$ equal to the renormalization scale
$\mu_R$, as assumed by the MMHT~2014 PDF.

All calculations are done in the MOM scheme. For comparison, we present results
for the $\phi$-averaged cross section $C_0$ in the $\overline{\rm MS}$ scheme,
as implemented in Eq.~(\ref{Cn_int}).
In the latter case, we choose natural values for $\mu_R$, {\it i.e.} 
$\mu_R = \mu_N \equiv \sqrt{|\vec k_H||\vec k_J|}$, and two different values of
the factorization scale, $(\mu_F)_{1,2} = |\vec k_{H,J}|$, depending on which of
the two vertices is considered. We checked that the effect of using natural
values also for $\mu_F$, {\it i.e.} $\mu_F = \mu_N$, is negligible with respect
to our two-value choice.

\subsection{Used tools and uncertainty estimation}
\label{numerics}

All numerical calculations were done using {\tt JETHAD}, a \textsc{Fortran}
code we recently developed, suited for the computation of cross sections and
related observables for two-body final-state processes, and offering also
support in the study of multi-body final-state reactions. 
In order to perform numerical integrations, we interfaced {\tt JETHAD} with
specific \textsc{CERN} program libraries~\cite{cernlib} and with {\tt Cuba}
library integrators~\cite{Cuba:2005,ConcCuba:2015}. 
We made extensive use of the CERNLIB routines {\tt Dadmul} and {\tt WGauss},
while the {\tt Cuba} ones were mainly used for crosschecks. 
A two-loop running coupling setup with $\alpha_s\left(M_Z\right)=0.11707$ and
five quark flavors was chosen.
It is known that potential sources of uncertainty could be due to the
particular PDF and FF parametrizations used. For this reason, we did
preliminary tests by using three different NLO PDF sets, expressly: 
MMHT~2014~\cite{Harland-Lang:2014zoa},  
CT~2014~\cite{Dulat:2015mca}
and NNPDF3.0~\cite{Ball:2014uwa},
and convolving them with the four following NLO FF routines: 
AKK~2008~\cite{Albino:2008fy}, 
DSS~2007~\cite{DSS}, HKNS~2007~\cite{Hirai:2007cx} and
NNFF1.0~\cite{Bertone:2017tyb}. 
All PDF sets and the NNFF1.0 FF parametrization were used via the Les Houches
Accord PDF Interface (LHAPDF) 6.2.1~\cite{Buckley:2014ana}.
Our tests have shown no significant discrepancy 
when different PDF sets are used in our kinematic range. 
In view of this result, in the final calculations 
we selected the MMHT~2014~PDF set, together with the FF interfaces mentioned
above. We do not show the results with DSS~2007 and NNFF1.0~FF routines, 
since they would be hardly distinguishable from those with 
the HKNS~2007 parametrization.

The most relevant uncertainty comes from the numerical 4-dimensional
integration over the two transverse momenta $|\vec k_{H,J}|$, 
the hadron rapidity $y_H$, and over $\nu$.  
Its effect was directly estimated by {\tt Dadmul} integration
routine~\cite{cernlib}.
The other three sources of uncertainty, which are respectively:
the one-dimensional integration over the parton fraction $x$
needed to perform the convolution between PDFs and FFs 
in the LO/NLO hadron impact factors,
the one-dimensional integration over the longitudinal momentum fraction
$\zeta$ in the NLO hadron/jet impact factor corrections,
and the upper cutoff in the numerical integrations over $|\vec k_{H,J}|$ and
$\nu$, are negligible with respect to the first one. For this reason the error
bands of all predictions presented in this work are just those given by
the {\tt Dadmul} routine.

\subsection{Discussion}
\label{discussion}

In Fig.~\ref{fig:C0_MSb_NS_CMS} we present our results at natural scales for
the $\phi$-averaged cross section $C_0$ at $\sqrt{s} = 7$ and 13 TeV in the
{\it CMS-jet} kinematic configuration. We can see that the NLO corrections
become larger and larger at increasing $Y$, an expected phenomenon in the
BFKL approach.

In Figs.~\ref{fig:Cn_MOM_BLM_CMS_7} and~\ref{fig:Cn_MOM_BLM_CMS_13},
predictions with the BLM scale optimization for $C_0$ and
several $R_{nm} \equiv C_n/C_m$ ratios with the jet tagged inside the CMS
detector are shown for $\sqrt{s} = 7$ and 13~TeV, respectively.
Here the benefit of the use of BLM optimization appears, since the LLA and
NLA predictions for $C_0$ are now comparable, a sign of stabilization of the
perturbative series. The trend of ratios of the form $R_{n0}$ is the standard
one and indicates increasing azimuthal decorrelation between the jet and the
hadron as $Y$ goes up, with the NLA predictions systematically above the
LLA ones, as it was also observed in Mueller-Navelet jets and in the
hadron-hadron case. The ratios $R_{21}$ and $R_{32}$ seem to be almost
insensitive to the NLO corrections.

Panels in Fig.~\ref{fig:Cn_MOM_BLM_CASTOR_13} show results with BLM scale
optimization for $C_0$ and several $R_{nm}$ ratios in the {\it CASTOR-jet}
configuration at $\sqrt{s} = 13$ TeV. They exhibit some new and, to
some extent, unexpected features: (i) the two parametrizations for the FFs
lead to clearly distinct predictions, (ii) $\langle \cos \phi\rangle$ exceeds
one at the smaller values for $Y$, a clearly unphysical effect. The
reason for these phenomena could reside in the fact that, the lower values
for $Y$ in the {\it CASTOR-jet} case are obtained for negative values of the
hadron rapidity, {\it i.e.} in final-state configurations where both
jet and hadron are backward.

Finally, in Fig.~\ref{fig:C0_comp_NLA_BLM_CMS} we compare the $\phi$-averaged
cross section $C_0$ in different NLA BFKL processes: Mueller-Navelet jet,
hadron-jet and hadron-hadron production, for $\mu_F = \mu_R^{\rm BLM}$, at 
$\sqrt{s} = 7$ and 13 TeV, and $Y \leq 7.1$ in the {\it CMS-jet} case.
The hadron-hadron cross section, with the kinematical cuts adopted,
dominates over the jet-jet one by an order of magnitude, with the hadron-jet
cross section lying, not surprisingly, in-between.

\section{Summary}

In this paper we have proposed a new candidate probe of BFKL dynamics
at the LHC in the process for the inclusive production of an identified
charged light hadron and a jet, separated by a large rapidity gap.

We have given some arguments that this process, though being a naive
hybridization of two already well studied ones, presents some own
characteristics which can make it worthy of consideration in future
analyses at the LHC.

In view of that, we have provided some theoretical predictions, with
next-to-leading accuracy, for the cross section averaged over the azimuthal
angle between the identified jet and hadron and for ratios of the azimuthal
coefficients.

The trends observed in the distributions over the rapidity interval between
the jet and the hadron are not different from the cases of Mueller-Navelet
jets and hadron-hadron, when the jet is detected by CMS, whereas some new
features have appeared when the jet is seen by CASTOR, which deserve further
investigation.

\section{Acknowledgments}

We thank C.~Royon and D.~Sunar~Cerci for fruitful discussions.
\\
F.G.C. acknowledges support from the Italian Foundation ``Angelo~della~Riccia''.
\\
D.I. thanks the Dipartimento di Fisica dell'U\-ni\-ver\-si\-t\`a della Calabria
and the Istituto Nazio\-na\-le di Fisica Nucleare (INFN), Gruppo collegato di
Cosenza, for the warm hospitality and the financial support.

\end{document}